\documentclass[apj]{emulateapj}

\newcounter{minirefcount}
\usepackage{amssymb}
\usepackage{amsmath}
\usepackage{wasysym}
\usepackage{lscape}
\usepackage{rotating}
\usepackage{apjfonts}
\newcommand{\mr}[2]{\refstepcounter{minirefcount}\label{#2}(\arabic{minirefcount}) #1}

\newcommand{\hi}{H~{\sc i}}
\newcommand{\ho}{H$_2$O}

\newcommand{\neii}{[Ne~{\sc ii}]}

\newcommand{\Ha}{H$\alpha$}

\newcommand{\Lsun}{L$_\odot$}
\newcommand{\Msun}{M$_\odot$}
\newcommand{\Lacc}{$L_{acc}$}
\newcommand{\Macc}{$\dot M_{acc}$}

\shorttitle{Probing stellar accretion with mid-infrared hydrogen lines}
\shortauthors{Rigliaco et al.}

\begin{document}
\title{Probing stellar accretion with mid-infrared hydrogen lines}  
\author{Elisabetta Rigliaco $^1$}
\affil{Department of Planetary Science, University of Arizona, 1629 E. University Blvd. Tucson, AZ 85719, USA}
\affil{Institute for Astronomy, ETH Zurich, Wolfgang-Pauli-Strasse 27, CH-8093 Zurich, Switzerland}
\email{$^1$rigliaco@lpl.arizona.edu, elisabetta.rigliaco@phys.ethz.ch}
\author{I. Pascucci}
\affil{Department of Planetary Science, University of Arizona, 1629 E. University Blvd. Tucson, AZ 85719, USA}
\author{G. Duchene}
\affil{Astronomy Department, University of California, Berkeley, Hearst Field Annex B-20, Berkeley CA 94720-3411, USA}
\affil{Universite de Grenoble Alpes, IPAG, F-38000 Grenoble, France}
\affil{CNRS, IPAG, F-38000 Grenoble, France}
\author{S. Edwards}
\affil{Five College Astronomy Department, Smith College, Northampton, MA 01063, USA}
\author{D.R. Ardila}
\affil{NASA Herschel Science Center, California Institute of Technology, MC 100-22, Pasadena, CA 91125, USA}
\affil{The Aerospace Corporation, M2-266, El Segundo, CA 90245, USA}
\author{C. Grady}
\affil{Eureka Scientific,2452 Delmer Street, Suite 100 Oakland, CA 94602-3017}
\affil{Exoplanets and Stellar Astrophysics Lab, Code 667, GoddardSpace Flight Center, Greenbelt, MD 20771}
\author{I. Mendigut\'ia}
\affil{School of Physics and Astronomy, University of Leeds, Woodhouse Lane, Leeds LS2 9JT, UK}
\author{B. Montesinos}
\affil{Departamento de Astrof\'isica, Centro de Astrobiolog\'ia, ESAC Campus, PO Box 78, 28691 Villanueva de la Ca\~nada, Madrid, Spain}
\author{G. D. Mulders}
\affil{Department of Planetary Science, University of Arizona, 1629 E. University Blvd. Tucson, AZ 85719, USA}
\author{J.R. Najita}
\affil{National Optical Astronomy Observatory, 950 North Cherry Avenue, Tucson, AZ 85719, USA}
\author{J. Carpenter}
\affil{Department of Astronomy, California Institute of Technology, MC 249-17, Pasadena, CA 91125, USA}
\author{E. Furlan}
\affil{Infrared Processing and Analysis Center, California Institute of Technology, 770 S. Wilson Ave., Pasadena, CA 91125, USA. }
\author{U. Gorti}
\affil{NASA Ames Research Center, Moffett Field, CA, USA}
\affil{SETI Institute, Mountain View, CA, USA}
\author{R. Meijerink}
\affil{Leiden Observatory, Leiden University, P.O. Box 9513, NL-2300 RA, Leiden, The Netherlands}
\author{M.R. Meyer}
\affil{Institute for Astronomy, ETH Zurich, Wolfgang-Pauli-Strasse 27, CH-8093 Zurich, Switzerland}

\begin{abstract}
In this paper we investigate the origin of the  mid-infrared (IR) hydrogen recombination lines for a sample of 114 disks in different evolutionary stages (full, transitional and debris disks) collected from the {\it Spitzer} archive. We focus on the two brighter {\hi} lines observed in the {\it Spitzer} spectra, the {\hi}(7-6) at 12.37$\mu$m and the {\hi}(9-7) at 11.32$\mu$m. We detect the {\hi}(7-6) line in 46 objects, and the {\hi}(9-7) in 11. We compare these lines with the other most common gas line detected in {\it Spitzer} spectra, the {\neii} at 12.81$\mu$m. 
We argue that it is unlikely that the {\hi} emission originates from the photoevaporating upper surface layers of the disk, as has been found for the {\neii} lines toward low-accreting stars. 
Using the {\hi}(9-7)/{\hi}(7-6) line ratios we find these gas lines are likely probing gas with hydrogen column densities of 10$^{10}$-10$^{11}$~cm$^{-3}$. 
The subsample of objects surrounded by full and transitional disks show a positive correlation  between the accretion luminosity and the {\hi} line luminosity. These two results suggest that the observed mid-IR {\hi} lines trace gas accreting onto the star in the same way as other hydrogen recombination lines at shorter wavelengths. A pure chromospheric origin of these lines can be excluded for the vast majority of full and transitional disks.
We report for the first time the detection of the {\hi}(7-6) line in eight young (< 20~Myr) debris disks. A pure chromospheric origin cannot be ruled out in these objects. If the {\hi}(7-6) line traces accretion in these older systems, as in the case of full and transitional disks, the strength of the emission implies accretion rates lower than 10$^{-10}$M$_{\odot}$/yr. 
We discuss some advantages of extending accretion indicators to longer wavelengths, and the next steps required to pin down the origin of mid-infrared hydrogen lines.

\end{abstract}
 

\section{INTRODUCTION}
\label{intro}

The analysis and characterization of circumstellar disks, which provide the raw material to form planets, are crucial for our understanding of disk dispersal and planet formation processes. 
Observations and modeling of the dust content in protoplanetary disks are nowadays relatively abundant (see Testi et al. 2014 for a review).  
The gaseous component of circumstellar disks is instead less well characterized (e.g., Dutrey et al. 2014 for a review) even if major advances have been made in the past few years thanks to space observatories such as {\it Spitzer} (e.g., Pontoppidan et al. 2014) and {\it Herschel} (e.g., Dent et al. 2013). 
Gas plays a pivotal role in the structure and evolution of disks representing the bulk of the disk mass (e.g., Alexander et al. 2014). It is then necessary to extend our knowledge of the disk gas content in order to improve our understanding of the natal environments of planets. 

The sensitivity of the InfraRed Spectrograph (IRS) on the {\it Spitzer} Space Telescope led  to the discovery of gas lines that can be used to characterize the immediate surroundings of young stellar objects (YSOs). 
For instance, the rich water spectrum of YSOs has been attributed to emission arising from warm ($\sim$600~K) layers of water vapor out to a few AUs from the central star (Carr \& Najita 2008, 2011, Salyk et al. 2008, Pontoppidan et al. 2010, Salyk et al. 2011, Banzatti et al. 2013). 

Among the wide forest of lines revealed by {\it Spitzer} the most common one is the singly ionized neon,  {\neii} at 12.81$\mu$m (e.g., Pascucci et al. 2007, Lahuis et al 2007, G\"{u}del et al. 2010). 
Different studies of this line show that objects with known jets/outflows have higher {\neii} luminosity than objects without (G\"{u}del et al. 2010, Baldovin-Saavedra et al. 2011). 
Moreover, follow up observations of this line with ground-based high-resolution spectrographs have shown that {\neii} is predominantly tracing an outflow in high-accreting stars, while its origin is from photoevaporative disk winds driven by  high-energy photons from the central star in lower accretors (Pascucci \& Sterzik 2009, van Boekel et al. 2009, Sacco et al. 2012, Baldovin-Saavedra et al. 2012). 

Other gas lines that are commonly used to characterize objects undergoing accretion are hydrogen recombination lines.
{\hi} lines from the near-ultraviolet (UV) to the near-infrared (IR) have been extensively used as indicators of the rate at which the matter from the inner circumstellar disk is accreting onto the star (e.g., Muzerolle et al. 2001, Calvet et al. 2004, Natta et al. 2006, Alcal\`a et al. 2014, Edwards et al. 2013 among many others). Recently, Salyk et al. (2013) using NIRSPEC at the Keck II telescope and CRIRES at the Very Large Telescope have introduced the {\hi} Pf$\beta$ line at 4.65$\mu$m as accretion diagnostic. This is to date the longest wavelength hydrogen line proposed as accretion indicator. 

As early as 2007, Pascucci et al. (2007) reported also the detection of {\hi} lines in the {\it Spitzer} spectrum of RX1852.3-3700. 
Subsequently, Ratzka et al. (2007) and Najita et al. (2010) reported hydrogen line detections in both the low- and high-resolution {\it Spitzer} spectrum of TWHya.  The analysis of these two objects did not constrain the origin of these lines and left several possibilities open: stellar chromosphere, hot disk atmosphere, accretion shock regions, or internal wind shock very close to the stellar surface (Hollenbach \& Gorti 2009).   More recently, Carr \& Najita (2011) reported the detection of the IR hydrogen lines in 10 T Tauri stars but did not discuss their origin. 

Here, we improve upon studies of gas lines from circumstellar disks by analyzing hydrogen lines observed in the {\it Spitzer}/IRS spectra of a large sample of objects. In particular we focus on the strongest of the Humphreys Series lines in the mid-infrared, the Hu$\alpha$ at 12.37$\mu$m (hereafter {\hi}(7-6)) and the {\hi}(9-7) line at 11.32$\mu$m. 
With the aim to investigate the origin of these mid-IR hydrogen lines, we compare their properties both with hydrogen lines observed at shorter wavelengths, mainly tracing accretion, and with the other most common gas line observed in the {\it Spitzer} spectra ({\neii}) which is instead an outflow or disk wind tracer depending on the mass accretion rate. 

The paper is organized as follows. In Section 2 the source selection and data reduction are explained. In Section 3 we describe the method used to account for the water contamination of the {\hi} lines, and how we measure the {\hi} and {\neii} line fluxes and luminosities. In Section 4 we discuss the origin of the {\hi} mid-IR lines, highlighting the interesting presence of {\hi} lines in a sample of debris disks. The summary of the results and an outlook are given in Section 5. 

\section{SAMPLE SELECTION AND DATA REDUCTION}
\label{section:obs}

Using the {\it Spitzer} Heritage Archive\footnote{http://sha.ipac.caltech.edu/applications/Spitzer/SHA/} we downloaded and re-reduced all IRS Short-High (SH) spectra of disks-bearing young stars acquired in the Standard Staring Mode and which have multiple on-source exposures (N-cycle $>$1) and sky observations. The latter two restrictions were applied to better identify and correct for bad pixels as summarized below. A list of the 114 objects identified in this way and with a signal-to-noise on the continuum greater than 5 is reported in Table~\ref{all_obj_log}.  
This table also contains the source coordinates, the Astronomical Observation Request number (AOR), the Identification Number (ID) of the {\it Spitzer} program the object belongs to, and the exposure time per frame times the number of exposures.
We also report in Table~\ref{tab_all_prop} the sources' properties, and in Table~\ref{flux_obj} the mid-IR line fluxes, errors and continuum levels around the three line of interest.  

\begin{longtable*}{l c c c c r}
\tablecaption{All objects reduced: Observing log\label{all_obj_log}}
\tablehead{\colhead{Name} & \colhead{RA} 	&\colhead{DEC}		&\colhead{AOR} 	& \colhead{Program ID}	&  \colhead{Time$\times$Ncycles}	}\\
\centering
\startdata
HD~377		&	  0h08m25.76s	&	    +6d37m00.3s          &  13462272	&	148	&   120 $\times$ 3
\\
HD~12039	&	  1h57m48.97s	&	    -21d54m05.2s         &  13461760	&	148	&   120 $\times$ 4
\\
HD~17925	&	  2h52m32.13s	&	    -12d46m11.0s         &   9780480	&	148	&   30 $\times$ 5
\\
HD~19668	&	  3h09m42.28s	&	    -9d34m46.4s          &  13462784	&	148	&   120 $\times$ 3
\\
LkHa~326		&	  3h30m44.01s 	&	   +30d32m47.0s         &  27063552	&	50641	&  30 $\times$ 12
\\
IC348-67		&	3h43m44.62s		&	+32d08m17.9s	&	22850816	&	40247	&  30 $\times$ 100
\\
IC348-31		&	3h44m18.16s		&	+32d04m57.0s	&	22849280	&	40247	&   30 $\times$ 16
\\
IC348-72		&	3h44m22.58s		&	+32d01m53.7s	&	22851840	&	40247	&   30 $\times$ 100
\\
IC348-68		&	3h44m28.51s		&	+31d59m54.1s	&	22851328	&	40247	&  30 $\times$ 100
\\
IC348-55		&	3h44m31.36s		&	+32d00m14.7s	&	22850304	&	40247	&   30 $\times$ 80
\\
IC348-2		&	3h44m35.36s		&	+32d10m04.6s	&	22847744	&	40247	&   30 $\times$ 4
\\
IC348-6		&	3h44m36.94s		&	+32d06m45.4s	&	22848256	&	40247	&   30 $\times$ 16
\\
IC348-37		&	3h44m37.99s		&	+32d03m29.8s	&	22849792	&	40247	&   30 $\times$ 30
\\
IC348-133	&	3h44m41.74s		&	+32d12m02.4s	&	22852352	&	40247	&  30 $\times$ 100
\\
IC348-21		&	3h44m56.15s		&	+32d09m15.5s	&	22848768	&	40247	&   30 $\times$ 30
\\
LkHa~330		&	  3h45m48.28s 	&     +32d24m11.9s         &  27063040		&	50641	&   30 $\times$ 7
\\
HD~25457	&	  4h02m36.74s	&	    -0d16m08.1s          &   9779712	&	148		&   6 $\times$ 4
\\
FM~Tau 		&	  4h14m13.58s 	&	   +28d12m49.3s         &  24403968	&	 30300	&   120 $\times$ 2
\\
FN~Tau		&	  4h14m14.59s 	&	   +28d27m58.0s         &  27062528	&	50641	&   30 $\times$ 6
\\
CW~Tau		&	  4h14m17.00s 	&	   +28d10m57.8s         &  27062272	&	50641	&   30 $\times$ 6
\\
FP~Tau		&	4h14m47.30s		&	+26d46m26.4s	&	26321408	&	50498	&   30 $\times$ 8 
\\
CX~Tau		&	  4h14m47.86s 	&	   +26d48m11.0s         &  23551744	&	40338	&   30 $\times$ 21
\\
CY~Tau		&	  4h17m33.72s 	&	   +28d20m46.8s         &  27061504	&	50641	&  30 $\times$ 14
\\
DD~Tau		&	4h18m31.13s		&	+28d16m28.9s 	&	26320384	&	50498	& 30 $\times$ 2
\\
BP~Tau		&	4h19m15.84s		&	+29d06m27.0s		& 14548224	&	20363	&  30 $\times$ 12
\\
DE~Tau		&	  4h21m55.67s 	&	   +27d55m06.2s         &  27060992	&	50641	&   30 $\times$ 10
\\
FT~Tau		&	  4h23m39.18s 	&	   +24d56m14.3s         &  27060224	&	50641	&   30 $\times$ 14
\\
04216+2603	&	  4h24m44.58s 	&	   +26d10m14.2s         &  27059968	&	50641	&  30 $\times$ 12
\\
IP~Tau 		&	  4h24m57.08s 	&	   +27d11m56.3s         &  24404224	&	30300	&   120 $\times$ 5
\\
DG~Tau		&	4h27m04.70s		&	+26d06m15.8s		& 14547968	&	20363	&	6 $\times$ 12 
\\
DH~Tau		&	4h29m41.57s		&	+26d32m58.2s	& 	26320128	&	50498	& 30 $\times$ 10 
\\
UX~TauA		&	  4h30m04.00s	&	+18d13m49.4s         &  18013952 &	30300 	&   30 $\times$ 30
\\
DK~Tau		&	4h30m44.28s		&	+26d01m24.6s 	&   14548736		&	20363	& 30 $\times$ 8
\\
HK~Tau		&	  4h31m50.58s 	&	   +24d24m17.8s         &  27059200	&	50641	&  30 $\times$ 14
\\
FZ~Tau		&	  4h32m31.77s 	&	   +24d20m02.6s         &  27058688	&	50641	&  30 $\times$ 6
\\
HN~Tau		&	  4h33m39.36s 	&	   +17d51m52.3s         &  27058176	&	50641	&   30 $\times$ 6
\\
DL~Tau		&	 4h33m39.08s  	&	  +25d20m38.1s         &  27058432	&	50641	&   30 $\times$ 6
\\
DM~Tau		&	  4h33m48.72s	&	   +18d10m10.0s         &  18014720	&	30300	&   30 $\times$ 30
\\
AA~Tau		&	  4h34m55.44s	&	  +24d28m53.4s 	     & 14551552	&	20363	&   30 $\times$ 12
\\
DN~Tau 		&	  4h35m27.37s 	&	   +24d14m58.9s         &  24404480	&	 30300	&   30 $\times$ 9
\\
DO~Tau		&	4h38m28.58s		&	+26d10m49.4s		& 14548480	&	20363	&	30 $\times$ 8
\\
LkCa~15		&	  4h39m17.80s 	&	   +22d21m03.5s         &  23551488	&	40338	&   30 $\times$ 21
\\
04385+2550	&	  4h41m38.82s 	&	   +25d56m27.0s         &  27057408	&	50641	&  30 $\times$ 6
\\
DP~Tau		&	  4h42m37.69s 	&	   +25d15m37.3s         &  27057152	&	50641	&   30 $\times$ 6
\\
DQ~Tau		&	4h46m53.06s		&	+17d00m00.4s	& 	26321152	&	50498	& 30 $\times$ 2 
\\
DR~Tau		&	  4h47m06.22s 	&	   +16d58m42.9s         &  27067136	&	50641	&   6 $\times$ 12
\\
DS~Tau		&	4h47m48.60s		&	+29d25m11.6s	& 26320896		&	50498	& 30 $\times$ 4 
\\
UY~Aur		&        4h51m47.38s	&	+30d47m13.9s 	     & 14551040	&	20363	&   6 $\times$ 10 
\\
St~34		&	  4h54m23.68s	&	    +17d09m53.5s         &  18016256	&	30300	&   30 $\times$ 40
\\
GM~Aur		&	  4h55m10.98s	&	    +30d21m59.5s         &  18015488	&	 30300	& 30 $\times$ 10
\\
SU~Aur		&	  4h55m59.38s 	&	   +30d34m01.5s         &  27066880	&	50641	&   6 $\times$ 10
\\
HD~32297		&	5h02m27.44s		&	+7d27m39.7s		&	25677568	&	50150	&   30 $\times$ 5
\\
V836~Tau		&	  5h03m06.60s 	&	   +25d23m19.7s         &  23552000	&	40338	& 30 $\times$ 21
\\
RW~Aur		&	5h07m49.56s		&	+30d24m05.0s	& 14547200		& 	20363	& 6 $\times$ 13
\\
HD~35850	&	  5h27m04.76s	&	    -11d54m03.5s         &   9777920	&	148	& 30 $\times$ 6
\\
HD~37484	&	  5h37m39.62s	&	    -28d37m34.7s         &   9780224	&	148	&  30 $\times$ 10
\\
NGC2068/ 2071-H9		&	5h46m11.86s	&	+0d32m25.9s	&	22853888	&	40247	& 30 $\times$ 100
\\
NGC2068/ 2071-S1		&	5h46m18.89s	&	-0d05m38.2s	&	22854912	&	40247	&  30 $\times$ 50
\\
NGC2068/ 2071-H13	&	5h46m22.44s	&	-0d08m52.6s	&	22854400	&	40247	&  30 $\times$ 80
\\
AO~Men		&	  6h18m28.24s	&	    -72d02m41.7s         &   5458688	&	148	& 120 $\times$ 2
\\
HD~61005	&	  7h35m47.47s	&	    -32d12m14.2s         &  13462528	&	148	& 120 $\times$ 3
\\
HD~69830	&	8h18m23.95s		&	-12d37m55.8s	&	12710656	&	41	&   30 $\times$ 2
\\
SX ~Cha		&	  10h55m59.76s	&	   -77d24m40.1s         &  27066624	&	50641	& 30 $\times$ 7
\\
TW~Cha	&	  10h56m30.52s	&	   -77d11m39.4s         &  27066368	&	50641	&   30 $\times$ 28
\\
SZ~Cha		&	10h58m16.77s	&	-77d17m17.1s	&	22846208	&	40247	&   30 $\times$ 4
\\
TW~Hya		&	  11h01m51.91s	&	   -34d42m17.0s         &  18017792	&	30300	&   30 $\times$ 2
\\
CS~Cha		&	  11h02m24.91s	&	   -77d33m35.7s         &  18021632	&	  30300	&  30 $\times$ 50
\\
Ced110-IRS2	&	  11h06m15.41s	&	   -77d21m56.8s         &  18022400	&	30300	&   30 $\times$ 10
\\
Ced110-IRS4	&	  11h06m46.58s	&	   -77d22m32.6s         &  18019328	&	30300	&   30 $\times$ 36
\\
Sz~18		&	  11h07m19.15s	&	   -76d03m04.8s         &  18020864	&	30300	&  30 $\times$ 100
\\
VW~Cha		&	  11h08m01.47s	&	   -77d42m28.7s         &  27066112	&	50641	&  30 $\times$ 6
\\
Sz~27		&	11h08m39.05s	&	-77d16m04.2s	&	22847232	&	40247	&   30 $\times$ 100
\\
VZ~Cha		&	  11h09m23.79s	&	   -76d23m20.8s         &  27065856	&	50641	&   30 $\times$ 8
\\
WX~Cha		&	  11h09m58.74s	&	   -77d37m09.0s         &  27065600	&	50641	&   30 $\times$ 12
\\
Hen~3-600A 	&	  11h10m27.53s	&	   -37d31m53.6s         &  18018560	&	30300	&   30 $\times$ 2
\\
XX~Cha		&	  11h11m39.66s	&	   -76d20m15.3s         &  27065344	&	50641	&  30 $\times$ 14
\\
RX~J1111.7-7620	&	  11h11m46.31s	&	   -76d20m09.1s         &   5451776	&	148	&   30 $\times$ 6
\\
CHX22		&	11h12m42.69s	&	-77d22m23.1s	&	22846720	&	40247	& 30 $\times$ 60
\\
1RXS~J121236.4-552037	&	  12h12m35.77s	&	   -55d20m27.4s         &   5460992	&	148	&   120 $\times$ 6
\\
1RXS~J122233.4-533347	&	  12h22m33.23s	&	   -53d33m48.9s         &  13463296	&	148	&    120 $\times$ 6
\\
TWA~10		&	  12h35m04.30s	&	   -41d36m39.0s         &  22934528	&	40338	&  30 $\times$ 10
\\
Sz~50		&	  13h00m55.38s	&	   -77d10m22.1s         &  27065088	&	50641	&   30 $\times$ 14
\\
1RXS~J130153.7-530446	&	  13h01m50.69s	&	   -53d04m58.1s         &  13462016	&	148	&    120 $\times$ 6
\\
1RXS~J132207.2-693812	(MPMus)&	  13h22m07.54s	&	   -69d38m12.1s         &   5451264	&	148	& 6 $\times$ 3
\\
1RXS~J133758.0-413448	&	  13h37m57.30s	&	   -41d34m42.0s         &   5460224	&	148	&   120 $\times$ 6
\\
HD~119269	&	  13h43m28.54s	&	   -54d36m43.5s         &   5461248	&	148	& 120 $\times$ 2
\\
HD~134319	&	  15h05m49.91s	&	   +64d02m50.0s         &   9779968	&	148	& 30 $\times$ 6
\\
V343Nor		&	  15h38m57.56s	&	   -57d42m27.3s         &   5458944	&	148	& 120 $\times$ 2
\\
HT~Lup		&	  15h45m12.87s	&	   -34d17m30.6s         &  27064832	&	50641	& 6 $\times$ 10
\\
GQ~Lup		&	  15h49m12.11s	&	   -35d39m04.8s         &  27064576	&	50641	&  30 $\times$ 8
\\
HD~141943	&	  15h53m27.30s	&	   -42d16m00.8s         &   5462528	&	148	&  120 $\times$ 4
\\
HD~142361	&	  15h54m59.86s	&	   -23d47m18.2s         &   5461504	&	148	&   120 $\times$ 12
\\
IM~Lup		&	  15h56m09.21s	&	   -37d56m05.9s         &  27064320	&	50641	& 30 $\times$ 6
\\
RU~Lup		&	  15h56m42.31s	&	   -37d49m15.5s         &  27064064	&	50641	&  30 $\times$ 5
\\
HD~143006	&	  15h58m36.93s	&	   -22d57m15.4s         &   9777152	&	148	&   6 $\times$ 8
\\
RX J1600.6-2159	&	  16h00m40.57s	&	   -22d00m32.3s         &  17145344	&	148 & 120 $\times$ 9
\\
EX~Lup		&	  16h03m05.49s	&	   -40d18m25.3s         &  27063808	&	50641	& 30 $\times$ 6
\\
RX~J1612.6-1859	&	  16h12m40.51s	&	   -18d59m28.3s         &   5457664	&	148	& 120 $\times$ 4
\\
RXJS~J161410.6-230542	&	  16h14m11.08s	&	   -23d05m36.2s         &   5453824	&	148	& 30 $\times$ 2
\\
1RXS~J161458.4-275013	&	  16h14m59.17s	&	   -27d50m23.0s         &  17144832	&	148	&  120 $\times$ 9
\\
DoAr~21 		&	  16h26m03.03s	&	   -24d23m36.4s         &  24403456	&	30300	& 30 $\times$ 6
\\
DoAr~24E		&	  16h26m23.42s	&	   -24d21m00.5s         &  27062784	&	50641	&  30 $\times$ 5
\\
DoAr~25		&	  16h26m23.69s	&	   -24d43m14.0s         &  24403712	&	30300	&  120 $\times$ 4
\\
SR~21 		&	  16h27m10.20s	&	   -24d19m16.0s         &  24403200	&	30300	& 6 $\times$ 10
\\
Haro~1-16	&	  16h31m33.46s	&	   -24d27m37.4s         &  27062016	&	50641	&   30 $\times$ 7
\\
RNO~90		&	  16h34m09.17s	&	   -15d48m16.8s         &  27061760	&	50641	&   6 $\times$ 9
\\
WaOph6		&	  16h48m45.63s	&	   -14d16m36.0s         &  27060736	&	50641	&  30 $\times$ 6
\\
EC~82		&	  18h29m56.90s	&	   +1d14m46.5s          &  27059712	&	50641	&   30 $\times$ 8
\\
RX~J1842.9-3532	&	  18h42m57.97s	&	   -35d32m42.7s         &   5451521	&	148	&   30 $\times$ 5
\\
RX~J1852.3-3700	&	  18h52m17.30s	&	   -37d00m11.9s         &   5452033	&	148	&  120 $\times$ 2
\\
AS~353A		&	19h20m30.98s	&	+11d01m54.8s	& 14547456		&	20363	& 30 $\times$ 8
\\
V1331 Cyg	& 	21h01m09.20s	&	+50d21m44.5s	&	14547712	&	20363	& 30 $\times$ 8
\\
HD~202917	&	  21h20m49.96s	&	   -53d02m03.2s         &   9778176	&	148	&  30 $\times$ 5
\\
HD~216803	&	  22h56m24.05s	&	   -31d33m56.1s         &   9777664	&	148	&   30 $\times$ 5
\\
\enddata
\end{longtable*}

\begin{figure}[h]
\centering
\includegraphics[scale=0.4, angle=0]{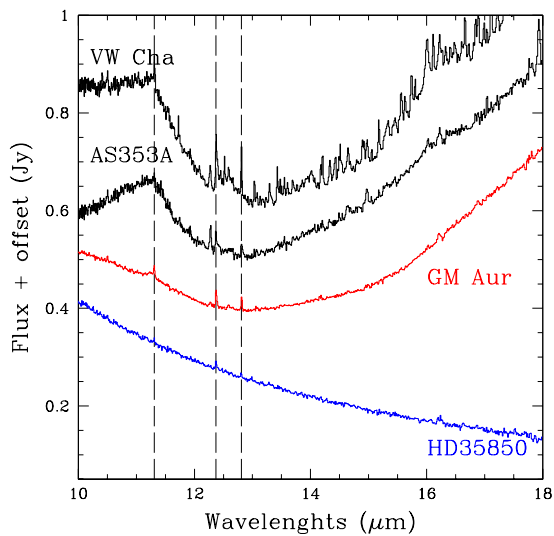}
\caption{{\it Spitzer}/IRS SH spectra of four disks. For plotting purposes offsets have been applied as follows: -0.1 Jy at VWCha (water-contaminated full disk), -0.6 Jy at AS~353A (water-free full disk), +0.3 Jy at GM~Aur (transitional disk) and 0.0 Jy at HD~35850 (debris disk). The vertical dashed lines show the location of the three lines mainly used through this paper: {\hi}(7-6) at 12.37$\mu$m, {\hi}(9-7) at 11.32$\mu$m and {\neii} at 12.81$\mu$m.  
 \label{ex_SH_spectra}}
\end{figure}

The IRS SH module uses a cross-dispersed echelle design and covers the spectral range 9.9-19.6 $\mu$m with a nominal resolving power of R$\sim$600 (Houck et al. 2004). The data reduction was performed starting from the {\it droopres} products following the procedure outlined in Pascucci et al. (2013). In brief, we computed the average of the on-source and sky exposures per nod position and rejected bad pixels using the noise statistics from multiple cycles. We then created background subtracted images and   
corrected for the additional known bad pixels reported in the IRS BCD mask. The one dimensional spectrum was extracted from the background-subtracted pixel-corrected images using the full aperture extraction routine in SMART (Higdon et al. 2004). We applied the same procedure to our sources and to the {\it Spitzer} flux calibrator $\xi$~Dra. From the reduction of 10 different exposures of $\xi$~Dra and the known
stellar model atmosphere we created two one-dimensional spectral
response functions (one at each nod position). Finally, we divided the source  spectra by the spectral response function. A compilation of four {\it Spitzer}/IRS SH spectra analyzed in this paper is shown in Figure~\ref{ex_SH_spectra}.

The calibrated spectra of 54 sources belonging to the program IDs 40247, 20363, 148 and 41 can be downloaded from the IRSA Spitzer Science Center at {\it http://irsa.ipac.caltech.edu/data/SPITZER/Disks\_SH\_spectra}. 

\begin{figure}[h]
\centering
\includegraphics[scale=0.4, angle=0]{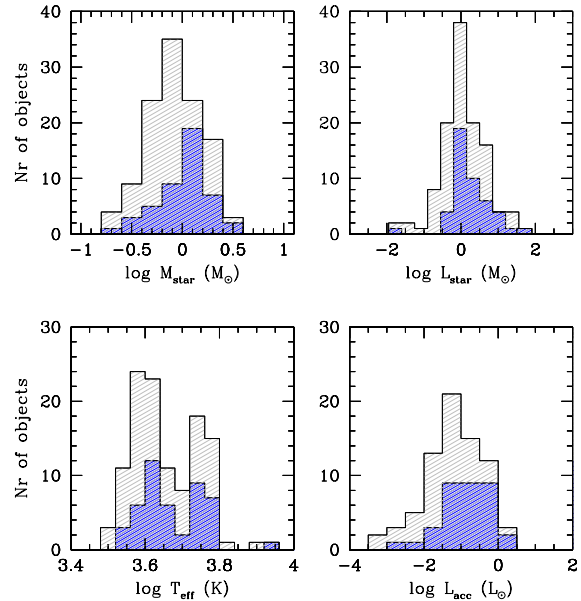}
\caption{Distribution of mass, luminosity, effective temperature and accretion luminosity for all the objects in our sample (gray), and only the objects where {\hi}(7-6) line has been detected (blue). 
 \label{isto_stellar_prop}}
\end{figure}

The total sample includes objects classified according to their general disk properties.  We use the generic term Full Disks (F) for those disks with strong excess emission above photospheric levels at infrared wavelengths, starting in the near-IR. 
Transitional Disks (T) are broadly defined as disks with reduced near- and mid-IR excess fluxes but excess emission comparable to full disks beyond $\sim$20$\mu$m, indicative of a depleted inner region. Finally, Debris Disks (D) are defined in this paper as having pure photosphere emission out to about 10\,\micron{} but can have relatively small excess emission in the long portion of the mid-IR spectrum. 
This disk classification collected from the literature is based on the dust properties and the continuum spectral energy distributions, and it is mainly taken from Furlan et al. (2006, 2011), Espaillat et al. (2012), Silverstone et al. (2006), Chen et al. (2014), as reported in Table~\ref{tab_all_prop}. 
Most of the debris disks included in our study belong to the recent catalog published by Chen et al. (2014). 
Full and transitional disks are primordial, gas-rich disks, while debris disks are second-generation disks, dominated by dust. 
In total our sample includes 68 full, 21 transitional, and 25 debris disks. 
In Figure~\ref{isto_stellar_prop} we show the distribution in mass, luminosity, effective temperature and accretion luminosity (see Sect.~\ref{accr_prop_sect}) for the total sample, and for the objects where the {\hi}(7-6) line is detected. 
A K-S two sample test of the stellar properties shows the two dataset have most likely the same parent distribution.

\section{ANALYSIS AND IMMEDIATE RESULTS}
In the following, we summarize the main properties of our sample (Sect.~\ref{sect_stellar_prop}), compute accretion luminosities for each object in  a self consistent manner (Sect.~\ref{accr_prop_sect}), and describe in detail how we calculate {\neii} and {\hi} fluxes taking into account any possible contamination from water lines (Sects.~\ref{ne_par} and \ref{hi_par}). The last two sections (Sects.~\ref{det_rates} and \ref{cont_sect}) discuss detections and correlations between line fluxes and adjacent continuum emission.

\subsection{Subsample of {\hi} and {\neii} sources: star and disk properties}
\label{sect_stellar_prop}

The properties of the 114 objects are summarized in Table~\ref{tab_all_prop}.    
For each object we provide the stellar association it belongs to, the distance, spectral type (SpTy), visual extinction (A$_V$) as well as the effective temperature (T$_{\rm eff}$) obtained using the T$_{\rm eff}$-SpTy correlation of Kenyon \& Hartmann (1995) and Luhman et al. (2003).  
H$\alpha$ equivalent widths and stellar luminosities (L$_{star}$) have been collected from the literature using the references reported in  Table~\ref{tab_all_prop}, while the mass (M$_{star}$), and radius (R$_{star}$) were derived from the HR diagram and the Siess et al. (2000) evolutionary tracks. 

For multiple star systems 
the name we report in Table~\ref{tab_all_prop} is that of the primary star, which is the one with the earliest spectral type. However, in the table caption we list the separation of any other stellar companion within 10$''$ of the primary star. 
Given that the IRS/SH slit width is 4.7$^{\prime\prime}$, there are 15 objects with companions closer than 2.5$''$ that can contaminate the extracted IRS spectra. However, McCabe et al. (2006) show that only in 25\% of the binary systems the secondary flux becomes larger than the primary flux at wavelengths $>$2\,\micron. Based on this, we assign the Spitzer spectrum always to the primary star in the 15 unresolved multiple systems.

We emphasize that the sample collected is biased in several ways: i) the number of objects with sky observations and multiple on-source exposures (see Sect.~\ref{section:obs}), ii) we only consider objects where the continuum has been detected and the signal-to-noise ratio of continuum level is higher than 5.  Hence, we do not attempt any statistical study based on this sample.

\subsection{Accretion Properties}
\label{accr_prop_sect}

To test whether mid-IR {\hi} lines trace the accretion shock region and/or the accretion flow, we compare {\hi} line luminosities with stellar accretion luminosities (\Lacc). One of the most direct methods to measure \Lacc\ is through modeling of the UV excess continuum emission produced by the accretion shock (e.g. Gullbring et al. 1998). However, only 17 objects in our sample have such measurements (AA~Tau, CS~Cha, DE~Tau, DK~Tau, DN~Tau, DM~Tau, DR~Tau, GM~Aur, HN~Tau, LkCa~15, RX~132207.2-693812, TW~Hya, SZ~Cha, Sz~27, LkHa~330, RX~J1852.3-3700 and RX~J1842.9-3532, see Ingleby et al. 2013 and Manara et al. 2014).  
To measure accretion luminosities in a consistent way for all the objects in our sample we will use the  \Ha\ line luminosity. 
Even if this method allows an homogeneous estimate of the accretion luminosity, we stress that the hydrogen lines in general, and the H$\alpha$ line in particular, have several contributing mechanisms: accretion, stellar winds, jets, chromosphere (e.g. Hartmann et al. 1990, Calvet \& Hartmann 1992). 
While this line is known to be an imperfect tracer of accretion in young stars,
Rigliaco et al. (2012), among others, have shown that \Lacc\ obtained from the UV excess emission and via indirect methods (e.g., \Ha\ line luminosity) agree to within an order of magnitude.

To measure the accretion luminosity we first compute the  \Ha\ luminosity from the \Ha\ equivalent width, the {\it R-} and/or {\it I-}band magnitude, and the extinction  (A$_V$) for each object (e.g., Hartigan et al. 1995). The data were dereddened using Mathis's reddening law for R$_{\rm V}$ =3.1 (Mathis 1990).  Next, we convert  the \Ha\ luminosity into  \Lacc\ using the relationship found by Rigliaco et al. (2012).
Although this method provides self-consistent values for \Lacc\, we are aware that the not-simultaneous observation of \Ha, photometry, and A$_V$ might introduce additional uncertainties in our estimates of \Lacc. In Sect.~\ref{HIasACCR} we describe how we account for these sources of uncertainty.


\subsection{{\neii} Line Fluxes and Luminosities}   
\label{ne_par}
 
 Water is largely detected in protoplanetary disks (e.g., Pontoppidan et al. 2010, Riviere-Marichalar et al. 2012), and its emission can contaminate other emission lines, as shown in the following sections. This is not the case for the {\neii} line, that is quite isolated from other {\ho} transitions. For this water-isolated and spectrally unresolved gas line we compute the flux by fitting a Gaussian profile to the observed line.  We define the continuum level around the line as the mean value obtained sampling a 0.1$\mu$m wide region shortward and longward of the transition, and the uncertainty on the measured flux is the rms in these regions. The line flux and the associated 1$\sigma$ uncertainty are reported in  Table~\ref{flux_obj} for each detection. In the same table we provide 3$\sigma$ upper limits to the flux when {\neii} lines are not detected. Line luminosities are then computed using the distances reported in  Table~\ref{tab_all_prop}. 

\begin{figure}[h]
\centering
\includegraphics[scale=0.4]{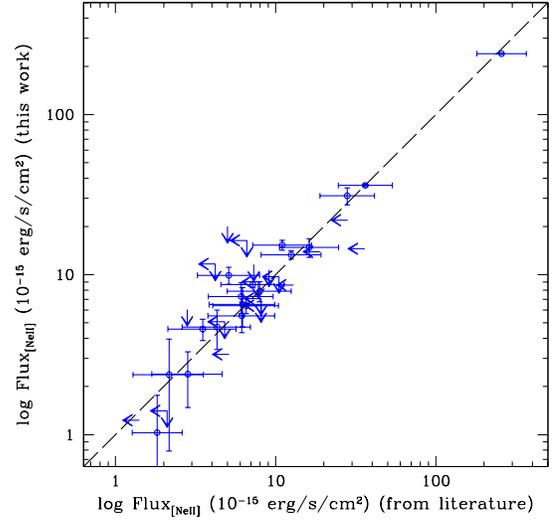}
\caption{{\small Comparison between the {\neii} line fluxes measured following the data reduction chain explained in Sect.~\ref{section:obs} and the line fluxes obtained from the literature using different data reduction techniques.
}
\label{neii_comparison}}
\end{figure}

We have compared the measured {\neii} line fluxes (as reported in Table~\ref{flux_obj}) with those reported by other authors using either different data reduction techniques (Pascucci et al. 2007, Lahuis et al. 2007, Espaillat et al. 2013),  and/or including any possible unresolved contribution from other species ({\ho} and OH, Carr \& Najita 2011). We find that the measured line fluxes and upper limits are in agreement within the errors (see Fig.~\ref{neii_comparison}). 

\subsection{{\hi} Line Fluxes and Luminosities}   
\label{hi_par}

The method used to compute the {\neii} line fluxes and luminosities could be in principle applied to the {\hi} lines. However, because the {\hi}(7-6) and {\hi}(9-7) transitions are very close to water lines we must subtract any possible contamination from water before measuring line fluxes.
We proceed as follows. 
In Sect.~\ref{sect_water1} we identify the objects that lack detectable water emission (hereafter, 'water-free' disks\footnote{We note that these objects might possibly show water contribution in higher S/N or higher resolution spectra.}) and the objects where water is detected in their spectra ('water-contaminated' disks). 
For the water-free disks  we compute line fluxes as in Sect.~\ref{ne_par}. 
For the water-contaminated disks we discuss a procedure to subtract water emission from the {\hi} lines (Sect.~\ref{sect_water2}) and then we measure the residual {\hi} fluxes after water subtraction.  

\subsubsection{Water-contaminated vs water-free disks}
\label{sect_water1}
To search for water emission in the source spectra we used the two water line complexes at 15.17$\mu$m and 17.22$\mu$m identified by Pontoppidan et al. (2010). These lines were selected because they are strong and more isolated than most other {\ho} lines. 
Following Pontoppidan et al. (2010), we define as water contaminated those spectra in which both the {\ho} lines are detected above 3$\sigma$.  We do not find any case where one of the two lines is detected above 3$\sigma$, and the other one is not detected. 

\begin{figure}[h]
\centering
\includegraphics[angle=-90,scale=0.33]{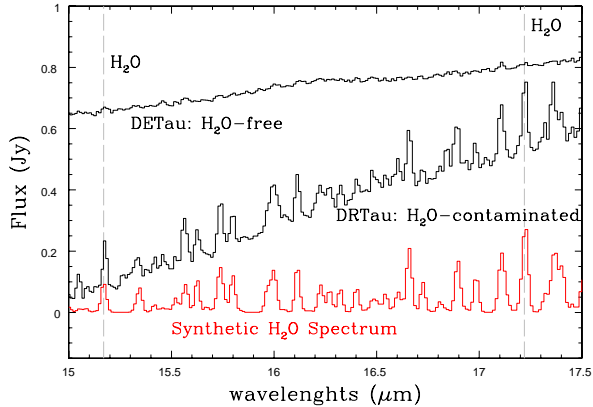}
\caption{{\small Examples of water contaminated (DR Tau), water-free (DE Tau) and synthetic {\ho} spectrum in the wavelength region around the two isolated {\ho} lines at 15.17 and 17.22 $\mu$m. The target spectra are shown in black, and the synthetic water spectrum is  shown in red.  
}
\label{figwater}}
\end{figure}

Among the 114 disks in our sample, 33 are water-contaminated. 
The remaining objects have water-free spectra.   
We show in Figure~\ref{figwater} a water-contaminated spectrum (DR~Tau), and a water-free spectrum (DE~Tau) in the region around the 15.17$\mu$m and 17.22$\mu$m water lines. For comparison we also show a synthetic water spectrum modeled assuming a gas temperature of 600~K and a water column density of 10$^{18}$~cm$^{-2}$ (see next Section). 
Line fluxes, together with the uncertainties and continuum level next to the lines are reported in Table~\ref{flux_obj}.

\subsubsection{Correcting for water emission}
\label{sect_water2}

Water emission lines have been detected towards many protoplanetary disks using the {\it Spitzer}/IRS spectrograph (Pontoppidan et al. 2014 for a review). 
Previous studies have shown that a simple model, which assumes a plane-parallel slab geometry and gas in local thermal equilibrium (LTE), can reproduce relatively well the fluxes of the observed {\ho} lines (e.g., Carr \& Najita~2011). However, we are aware that these assumptions ignore the non-LTE effects and the complex geometric structure of disks (see Meijerink et al. 2009). 
We use this simple model as developed in Pascucci et al.~(2013) to subtract any water contamination from the {\hi} lines. The specific LTE model takes the molecular parameters from the HITRAN 2008 database (Rothman et al. 2009) and requires only 4 inputs: the gas temperature, the water column density, the radius of the emitting area, and the line width (which we assume to be the thermal line width, i.e., variable with temperature).  
We created a grid of synthetic spectra using five different gas temperatures (T=[400, 500, 600, 700, 800]) and water column densities (N$_{col}$=[10$^{17}$, 5$\times$10$^{17}$, 10$^{18}$, 5$\times$10$^{18}$, 10$^{19}$~cm$^{-2}$]) and assumed the same projected emitting area of radius 1~AU. The chosen inputs span the range of parameters inferred for the warm disk atmosphere of T~Tauri stars (Carr \& Najita 2011, Salyk et al.~2011). 

\begin{figure}[h]
\centering
\includegraphics[angle=-90,scale=0.3]{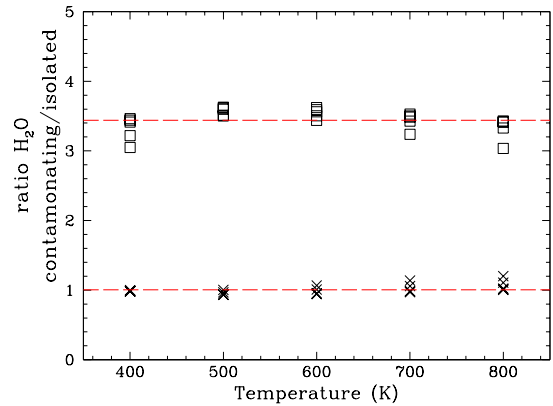}
\caption{{\small {\ho} contaminating/isolated line ratios: crosses for {\ho}(11.32$\mu$m)/{\ho}(11.97$\mu$m) and squares for {\ho}(12.39$\mu$m)/{\ho}(13.13$\mu$m). The ratios are computed for all the temperatures and column densities explored with the model grid. At each temperature there are 5 ratios, one for each value of the column density. 
}
\label{fig_ratio}}
\end{figure}

The water lines contaminating the hydrogen lines are centered at $\sim$12.39$\mu$m and $\sim$11.31$\mu$m for the {\hi}(7-6) and {\hi}(9-7) transition, respectively.  In the following we will refer to these lines as ``{\ho}-contaminating''. 
To subtract off the water contribution to the observed {\hi} lines we have searched the synthetic spectra for {\it isolated} {\ho} lines with similar properties as the {\ho}-contaminating ones. 
The two water lines that we have identified will  be called in the following ``{\ho}-isolated'' lines and have the following characteristics: they have similar lower energies (E$_{low}$) and Einstein A-coefficients (A$_{ul}$) (within 15\%  at most) as the {\ho}-contaminating lines; {\ho}-isolated line is within 1\micron\ from the corresponding contaminating line, so that the continuum level does not substantially change. 
Following these criteria we find that the {\ho}-isolated line at 13.13$\mu$m is the best match for the {\ho}-contaminating line at 12.39$\mu$m, while the {\ho}-isolated at 11.96$\mu$m has very similar properties as the {\ho}-contaminating at 11.31$\mu$m (see Table~\ref{line_prop}).  
We show in Fig.~\ref{fig_ratio} the ratio between the {\ho} contaminating/isolated line fluxes for all the temperature and column density of the synthetic spectra. The ratios are quite constant among all the synthetic spectra, with {\ho}(12.39$\mu$m)/{\ho}(13.13$\mu$m)$\sim$3.5 and {\ho}(11.32$\mu$m)/{\ho}(11.97$\mu$m)$\sim$1.0, and vary  within a factor 1.5 at most over our grid of  synthetic spectra.

Once we identified the best water proxies, we searched all the spectra for {\ho}-isolated lines. In the SH-{\it Spitzer}/IRS spectra none of the 83 water-free spectra shows these lines, as expected because the {\ho} lines become weaker at shorter wavelengths (e.g., Salyk et al. 2011). 
In the 33 water-contaminated disks, we subtracted the water contribution from the {\hi} line fluxes as follows: if the {\ho}-isolated line is detected above the 3$\sigma$ level, we use it and its ratio to the nearby {\ho}-isolated line to compute the {\ho}-contaminating flux (as {\ho}(12.39$\mu$m)=3.5$\times${\ho}(13.13$\mu$m) and {\ho}(11.32$\mu$m)={\ho}(11.97$\mu$m),  respectively for {\hi}(7-6) and {\hi}(9-7)), and then subtract it from the measured HI line flux. If the {\ho}-isolated is not detected, we subtract its 3$\sigma$ upper limit from the observed {\hi} flux. 
After subtracting the water contamination from the observed line flux, we obtain the actual hydrogen line flux.
The final line fluxes after water subtraction and errors are reported in Table~\ref{flux_obj}. We also show in parenthesis the total flux observed before subtracting the water contribution in the objects where the {\hi} lines are detected.  In the case of {\hi} non-detections we provide the 3$\sigma$ upper limit to the flux. 
Any other water contribution, if present, is within the 10\% absolute flux uncertainty. 
As for the {\neii}, we have compared our {\hi}(7-6) line flux measurements with the few already reported in the literature (Pascucci et al. 2007, Carr \& Najita 2011). We find that the reported values are in agreement within the errors and the absolute flux uncertainty.

\begin{deluxetable}{l c c c}
\tablecaption{{\ho} Line Properties  \label{line_prop}}
\tablehead{\colhead{Line} & \colhead{Wavelength} & \colhead{$A_{ul}$}  & \colhead{$E_{low}$} \\
	& $(\mu$m) 	& (s$^{-1}$)	& (cm$^{-1}$)} 
\tabletypesize{\footnotesize}
\startdata
H$_2$O-contaminating 	&	11.3241     &  	1.576	&	2746.02	\\
H$_2$O-contaminating 	&	12.3962	&	7.665 	&	3211.21 	\\
H$_2$O-isolated 	&	11.9681     &	1.281	&      2629.33	\\
H$_2$O-isolated 	&	13.1325     & 	6.788      	&	3244.60	\\
 \enddata
\end{deluxetable}

\subsection{Line Detections Summary}
\label{det_rates}

We detect the {\neii} line at 12.81\,\micron{} in 49 objects in our sample. Among the 67 objects where the {\neii} line is not detected, 37 are full disks, 4 are transitional disks and 26 are debris disks. One debris disk has a {\neii} line detection (Rigliaco et al. in preparation). 
After correcting for water contamination, we identify 46 objects with an {\hi}(7-6) detection above 3$\sigma$, 11 of them also show the weaker {\hi}(9-7) line. 
In 6 spectra we detect all three lines of interest: the {\neii}, the {\hi}(7-6), and the {\hi}(9-7) line.    
Line detections are summarized in Table~\ref{tab_number}.
We note that all objects where water emission is detected are full disks and none of the transitional or debris disk spectra have detectable water emission. Thus, expanding upon previous results (e.g., Pontoppidan et al. 2010), our analysis shows that the absence of water lines in the SH module is typical to more evolved transitional and debris disks, while full disks can either have a water rich spectrum or not. 

Interestingly, we detect {\hi}(7-6) in 8 young (<20~Myrs) debris disks. We will discuss these objects in Sect.~\ref{section_DD}.  

\begin{deluxetable}{l c | c c c}
\centering
\tablecaption{Line Detections  \label{tab_number}}
\tablehead{\colhead{Detected Lines} & \colhead{Total number} & \multicolumn{3}{c}{{Relative \%}} \\
&  of objects & F & T  & D }
\startdata
{\neii} & 49 & 67 & 31 & 2\\	
{\hi}(7-6) & 46 & 57 & 26 & 17 \\	
{\hi}(9-7) & 11 & 55 & 27 & 18\\	
\hline
{\neii}+{\hi}(7-6) & 24 & 50 & 46 & 4 \\	
{\hi}(7-6)+{\hi}(9-7) & 11 & 55 & 27 & 18\\	
{\neii}+{\hi}(7-6)+{\hi}(9-7) & 6 & 50 & 50 & 0\\	
\tablecomments{ The second column shows the total number of objects where the lines listed in Column 1 have been detected. The 3rd, 4th and 5th columns report the relative percent of these objects in full, transitional and debris disks respectively.
}
\end{deluxetable}

\subsection{{\hi} lines versus continuum luminosity}
\label{cont_sect}
\begin{figure}
\centering
\includegraphics[angle=-90,scale=0.33]{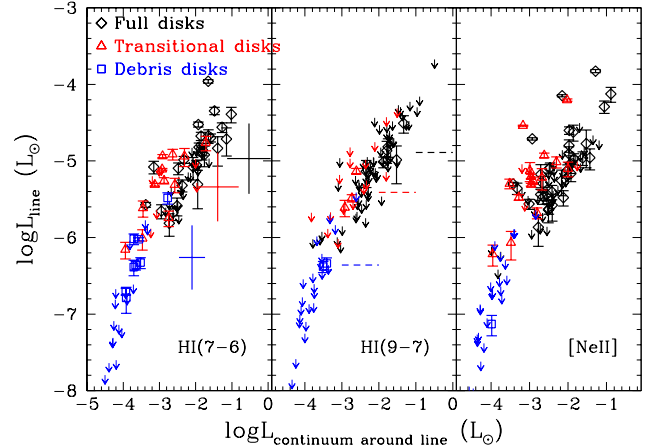}	
\caption{{\small{\hi}(7-6), {\hi}(9-7) and {\neii} line luminosity versus continuum luminosity around the lines. Black-diamonds represent objects surrounded by a full disk, red-triangles are for transitional disks, blue-squares for debris disks. Downward pointing arrows are upper limits. In the left panel solid lines represent the mean value of line and continuum luminosities with the 1$\sigma$ uncertainty for the objects where both line and continuum are detected. In all cases, the mean value for the x-axis has been shifted by +1.5 for clarity. In the middle panel we show the mean in the line luminosity as dashed lines, given the large fraction of non-detections. We note again that the measure of the means is only based on the detections, and for the {\hi}(9-7) lines the upper limits are more numerous than the detections. 
}
 \label{H76_cont}}
\end{figure}

Once we accounted for the water contribution and measured the {\hi}(7-6) and {\hi}(9-7) line luminosities, we investigated their behavior as a function of the continuum luminosity around the lines (Fig.~\ref{H76_cont}).  
As also seen in other IR lines (H$_2$S(2), [Fe~{\sc i}] at 24$\mu$m, {\neii}, e.g., Lahuis et al. 2007), there is a correlation between the {\hi} mid-IR lines and the continuum flux level next to the lines. This correlation could be influenced, but it is not exclusively driven, by the more favorable contrast between line and continuum luminosities where the continuum is lower (as demonstrated by finding the same increasing trend between the line to continuum vs continuum flux). 
We have also investigated this correlation for different disk morphologies and find that debris disks have lower {\hi} line luminosities and continuum luminosities than transitional and full disks. In order to understand whether these differences are statistically significant we have measured the mean and standard deviation of line and continuum luminosities for the three different disk morphologies (shown as solid lines in Fig.~\ref{H76_cont}). From this analysis we conclude that, even if transitional disks appear having lower {\hi} and continuum luminosities than full disks, their difference is within 1$\sigma$ uncertainty, and thus not statistically significant. 
On the other hand, debris disks show consistently lower luminosities, both in the continuum and in the lines.

\section{Discussion}

\subsection{Mid-IR {\hi} lines: disk tracers?}
\label{disk_tracers_sect}

In the following we compare the {\neii} line fluxes obtained from our reduction of the {\it Spitzer} spectra with those of the {\hi} lines to test if these two lines could trace the same region in/around the disk. In particular, we will test if {\neii} and {\hi} lines may trace the disk surface. 
As mentioned in Sect.~\ref{intro}, {\neii} lines have been studied both with {\it Spitzer}  and with ground-based spectrographs at higher spectral resolution. These studies have shown that the {\neii} line mostly traces jets/outflows in high-accreting stars and the hot disk surface in lower accretors (\Macc $\lesssim 5\times10^{-8} M_{\odot}$/yr, e.g. G\"{u}del et al. 2010, Sacco et al. 2012). 
Unfortunately, similar ground-based followup studies have not yet been carried out for the mid-IR {\hi} lines, hence a direct comparison of high-resolution spectra cannot be done.
 The vast majority of the objects in our sample (namely, 85\%) for which we have computed \Lacc\ (and \Macc) can be classified as ``low-accretors'' according to the threshold value reported above (G\"{u}del et al. 2010). Hence, for most of our sources a detection of {\neii} in the {\it Spitzer} spectrum is most likely due to emission from the hot disk surface.

\begin{figure}
\centering
\includegraphics[angle=-90,scale=0.33]{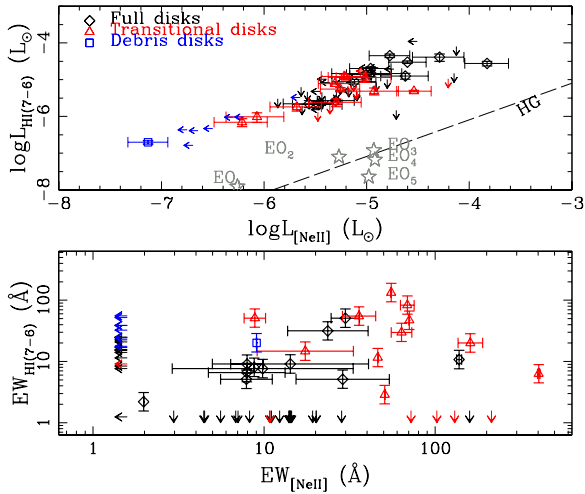}
\caption{{\small Top: {\hi}(7-6) versus {\neii} line luminosity. We also show the prediction of these line luminosities for different theoretical models. The black-dashed line (labeled as HG) shows the {\hi}(7-6)/{\neii} line relation for a disk irradiated by EUV photons (Hollenbach \& Gorti 2009). The gray stars refer to X-ray irradiated primordial disk with logL$_X$=29.3 (EO$_1$), logL$_X$=30.3 (EO$_2$), and transitional disks  illuminated by logL$_X$=30.3 erg/s with inner hole 8.3 AU (EO$_3$), 14.2 AU (EO$_4$) and 30.5 AU (EO$_5$), as measured by Ercolano \& Owen (2010).   Bottom: comparison between the equivalent widths of the two lines. }
 \label{LH76_LNe_comparison}}
\end{figure}

The left panel of Figure~\ref{LH76_LNe_comparison} shows a log-log plot of the {\hi}(7-6) line luminosity at 12.37$\mu$m versus the {\neii} line luminosity at 12.81$\mu$m . The luminosities of the two lines, when they are both detected, appear positively correlated, with {\it Pearson} correlation coefficient of $\sim$0.9, and p-value$\sim$10$^{-6}$. The correlation becomes weaker when upper limits are considered. However, this correlation could be affected by the underlying correlation between the line luminosities and the continuum luminosities around the line (see Figure~\ref{H76_cont} and discussion in Sect~\ref{cont_sect}). To account for this possible effect  we plot in Fig.~\ref{LH76_LNe_comparison} (right panel) the equivalent widths of the two observed lines. The {\neii} and {\hi}(7-6) equivalent widths do not show any obvious correlation, and the {\it Pearson} correlation coefficient drops to $\sim$0.4 (weak correlation) and p-value$\sim$0.04. This result suggests that the positive relation between the {\hi}(7-6) and {\neii} line luminosity is caused by the correlation between the line and nearby continuum. 

To further test this finding, we compare the observed  {\hi} line luminosities with those predicted by models of disks irradiated by high-energy photons (Ercolano et al. 2008, Hollenbach \& Gorti 2009, Ercolano \& Owen 2010).  
In an EUV (13.6~eV$<h\nu<$0.1~keV) irradiated disk the  predicted L$_{\rm HI(7-6)}$/ L$_{\rm [NeII]}$ ratios 
are as small as $\sim$0.008 (Hollenbach \& Gorti 2009) because the radiative recombination rate of hydrogen is smaller than the electronic collisional excitation rate of {\neii}. The observed line ratios are instead  almost two orders of magnitude higher than the predicted ones (see Table~\ref{flux_obj}). 
A disk irradiated also by X-rays ($h\nu>$0.1keV) still produces an {\hi}(7-6) line luminosity more than one order of magnitude smaller than the ones observed (Ercolano et al. 2008, Ercolano \& Owen 2010). 
The comparison between the observed line luminosity and different theoretical models shows that it is  
unlikely that the observed HI(7-6) lines trace the same hot disk atmosphere that produces the {\neii} lines. 

In summary, both the lack of correlation between the {\neii} and the {\hi}(7-6) equivalent widths, and the comparison with model predictions suggest that the {\hi}(7-6) line is not tracing the same disk region as the {\neii} line, and that the {\hi} lines are unlikely disk gas tracers.

\subsection{Mid-IR {\hi} lines: accretion indicators?}
\label{HIasACCR}

Hydrogen recombination lines from the Balmer to the Paschen and Brackett series (in the UV, optical and near-IR wavelengths) have been extensively studied over the years, and have been convincingly found to trace infalling material onto the star, as demonstrated by their broad line profiles consistent with gas at almost free-fall velocity onto the star  (e.g., Alcal\'a et al. 2014). Nevertheless, these line profiles are also known to have other contributing mechanisms: stellar wind, jet, chromosphere. 
Recently, Salyk et al. (2013), using high-resolution spectroscopy extended the accretion tracers to the Pfund series lines, including the Pf$\beta$ hydrogen recombination line at 4.65$\mu$m ({\hi}(7-5)). 

To test  if the {\hi} mid-IR lines can be used as accretion indicators, in the rest of this section we focus on the subsample of full and transitional disks for which an estimate of the H$\alpha$ equivalent width is available in the literature (see Table~\ref{tab_all_prop}). 
We chose to use the {\hi}(7-6) line (the strongest among the {\hi} lines observed with the {\it Spitzer} SH module) to maximize the number of detected sources. Among the objects considered in this work, 73 have an estimate of the accretion luminosity from the H$\alpha$ line (34 of those have an {\hi}(7-6) detection).

\begin{figure}
\centering
\includegraphics[scale=0.3, angle=-90]{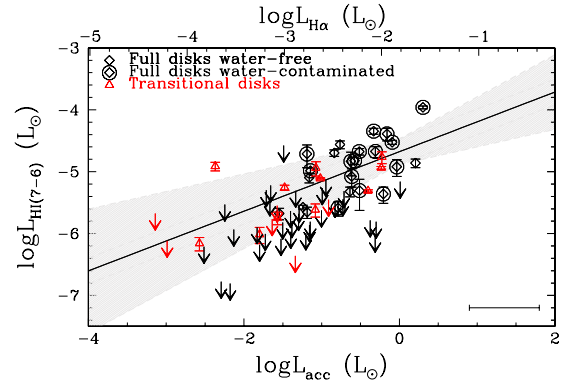}		
\caption{
{\small 
{\hi}(7-6) line luminosity versus \Ha\ line luminosity and L$_{acc}$. Circles are drawn around water-contaminated objects. The black-solid line shows the best linear regression line and the shaded-gray area represents the approximate 95\% confidence intervals (2$\sigma$) on the regression line. 
The typical error is shown in the bottom-right corner. 
}
 \label{H76_Lacc}}
\end{figure}

We show in  Fig.~\ref{H76_Lacc} the {\hi}(7-6) line luminosity as a function of the \Ha\ line luminosity and the accretion luminosity measured as described in Sect.~\ref{accr_prop_sect}. There is a linear correlation between logL$_{HI(7-6)}$--logL$_{acc}$.    
To estimate the best fit regression line for the objects with detection in both quantities, and to account for the measured errors in both the x- and y-axes we employe the Bayesian method as implemented in the IDL routine {\it linmix\_err} (Kelly 2007). The error on the y-axes is driven by the uncertainty on the observed line fluxes. The uncertainty in our estimate of \Lacc\ is given by the combination of several factors: the uncertainty on the H$\alpha$ equivalent width, the non-simultaneous observation of \Ha, photometry, and extinction, the error on the extinction value (A$_V$) and the errors on slope and intercept of the \Lacc-L$_{H\alpha}$ relationship. 
Following Costigan et al. (2012) we assume a mean amplitude of variation in the derived accretion luminosities of $\sim$0.37 dex, which includes H$\alpha$ equivalent width uncertainties, time variability, and the error in the \Lacc-L$_{H\alpha}$ relationship. We also consider an uncertainty on  A$_V$ of 1~mag (e.g. Edwards et al. 2013), which gives a final uncertainty on \Lacc\ $\sim$0.5~dex. 
We find the following best fit regression line between the {\hi}(7-6) line luminosity and the accretion luminosity: 
\begin{equation} 
logL_{HI(7-6)}/L_\odot = (0.48 \pm 0.09)\times logL_{acc}/L_\odot - (4.68 \pm 0.10)	
\label{acc_Lh_rel}
\end{equation}

We have checked that the log$L_{HI(7-6)}$--log$L_{acc}$ correlation is not driven by the distance and/or stellar luminosity dependence. 
We have also ascertained that the relationship between the luminosity of the {\hi}(7-6) line and \Lacc\ measured through UV-excess emission for only 15 objects (see Sect.~\ref{accr_prop_sect}) lies within the 95\% confidence intervals. 
We note that the uncertainty on the intercept we find in Eq.~(1) is similar to the those found using different {\hi} lines (e.g., Alcal\`a et al. 2014),  suggesting that the scatter around the fit is similar when using different {\hi} lines. On the other hand, the slope found in Eq.~(1) is shallower than the ones found for the {\hi} lines in the optical and near-IR wavelengths.  This suggests that the MIR lines are less sensitive to accretion luminosity variations.

Knowing \Lacc\ and the stellar properties we can also compute the mass accretion rate (\Macc) using the classical relation from Gullbring et al. (1998): 
\begin{equation}
\dot M_{acc} = \Big(1-\frac{R_{star}}{R_{in}}\Big)^{-1} \frac{L_{acc} R_{star}}{G M_{star}}
\label{macc_eq}
\end{equation}
where R$_{in}$ is the radius at which the accreting gaseous disk is truncated due to the stellar magnetosphere. 
In full disks, R$_{in}$ has been found ranging between 3R$_{star}$ and 10R$_{star}$ (Johns-Krull 2007), and usually R$_{in}$=5R$_{star}$ is assumed. In our sample we do not only consider full disks, but also transitional disks where the geometry of the gas disk around the star could be different. We do not speculate on how that could change for more evolved disks, and we will use the R$_{in}$=5R$_{star}$ for all the different disk morphologies. 

\begin{figure}
\centering
\includegraphics[angle=-90,scale=0.33]{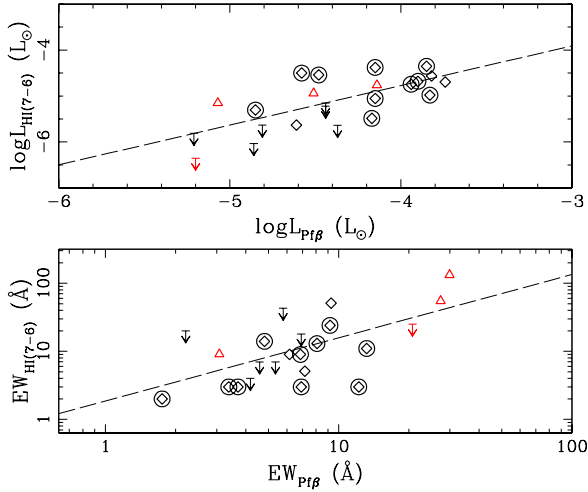}
\caption{{\small Top: {\hi}(7-6) versus Pf$\beta$ line luminosity. The black-dashed line hows the least square fit.   Bottom: comparison between the equivalent widths of the two lines. }
 \label{LH76_LPfB_comparison}}
\end{figure}

For the subsample of objects where the Pf$\beta$ line is detected  (Salyk et al. 2013) and the  {\hi}(7-6)  is detected and not water contaminated (this work) we find that their luminosities, as well as equivalent widths, are correlated with {\it Pearson} correlation coefficients $\sim$0.7 and $\sim$0.8, respectively,  and p-value $\sim$0.007 and $\sim$0.05, respectively (Fig.~\ref{LH76_LPfB_comparison}).  
While Salyk et al. 2013 report lower Pf$\beta$ luminosities and high EWs for 5 transitional disks with respect to full disks, we do not find any statistically significant difference in the {\hi}(7-6) lines in our much larger samples of transitional and full disks (see also Sect.~\ref{disk_tracers_sect}). We also note that these two samples have not statistically different \Lacc, as also reported in other studies where \Lacc are consistently derived (Keane et al. 2014, Fang et al. 2009).

The fact that the {\hi}(7-6) line correlates both with the H$\alpha$ and Pf$\beta$ lines suggests that 
the Humphreys $\alpha$ line at 12.37$\mu$m is likely an accretion indicator, as the hydrogen lines belonging to other series (H$\alpha$ (e.g., Muzerolle et al. 1998, Pa$\beta$ (e.g., Folha \& Emerson 2001), Br$\gamma$ (e.g., Natta et al. 2006), Pf$\beta$ (Salyk et al. 2013)). 

Speculating on the correlation reported in Eq.~(1), we can measure what the mass accretion rates would be if the {\hi}(7-6) lines in the debris disks spectra were accretion related. We find that debris disks with detected {\hi} lines span the lower range of mass accretion rate (covering the range between $\times$10$^{-10}$--2$\times$10$^{-12}$ M$_{\odot}$/yr), as would be expected by their more evolved  stage and the lack of gas in their disks.

\subsection{HI Line Ratios}

Hydrogen line ratios can be used as a diagnostic tool to constrain the physical conditions (density and temperature) of the emitting gas.
Among the sample of objects analyzed in this paper, in 46  the {\hi}(7-6) transition is detected above 3$\sigma$, and 11 of those show as well a detection of the {\hi}(9-7) transition. The detected {\hi}(9-7)/{\hi}(7-6) line ratios range between $\sim$0.4 and $\sim$1.1. The upper limits cover a similar range.
We can retrieve information on the physical properties of the emitting gas by comparing the observed ratios with those predicted by two different models: the Case B (Hummer \& Storey 1987, Storey \& Hummer 1995) and the Kwan and Fischer (Kwan \& Fischer 2011, KF) models. In these models the {\hi} line ratios ranges between $\sim$0.3--0.5 in the optically thin regime, and become higher (between $\sim$1.0--2.0) when the lines are optically thick.

\subsubsection{Case B models}
\label{caseBmod}

\begin{figure}[h]
\centering
\includegraphics[angle=-90,scale=0.3]{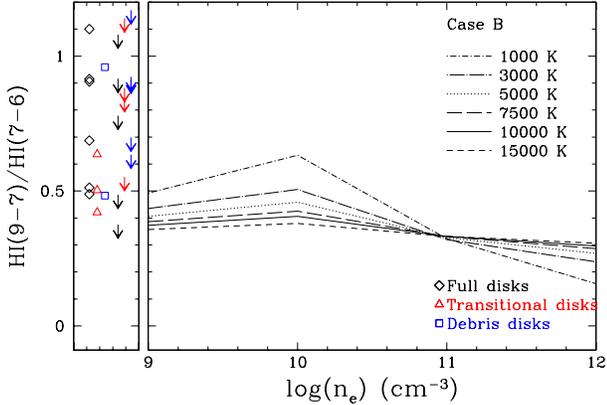}
\caption{{\small HI(9-7)/HI(7-6) line ratio versus the electron density. Case B models for different temperatures as labeled. The range of the observed ratios is shown along the left side, where the typical uncertainties on the observed ratios are also plotted. 
\vskip 0.8cm
}
\label{caseBmodels}}
\end{figure}

The Case B model for radiative ionization and recombination (Baker \& Menzel 1938) assumes that gas  is optically thick to Lyman series photons and optically thin to photons associated to all other transitions. 
This model has been largely used to retrieve the properties of gas accreting onto T~Tauri stars. 
However, results found considering individual objects and different hydrogen lines indicate very different physical properties of the accreting gas (Bary et al. 2008, Rigliaco et al. 2009, and Edwards et al. 2013 for a review). 
Using the interactive online server\footnote{http://cdsarc.u-strasbg.fr/viz-bin/Cat?VI/64} (Hummer \& Storey 1987, Storey \& Hummer 1995), we computed the predicted line ratios for a range of temperatures from 1000~K to 15,000~K and electron densities from 10$^{9}$--10$^{12}$cm$^{-3}$ (see Fig.~\ref{caseBmodels}). 

The predicted {\hi}(9-7)/{\hi(7-6)} line ratios are almost constant (between $\sim$0.3-0.4) for T$\geq$5000~K over the electron density range considered here. As the temperature decreases under 5000~K, the predicted ratio increases from $\sim$0.3 to $\sim$0.7 until electron densities $n_e \lesssim 10^{10}$cm$^{-3}$, and then decreases to $\sim$0.2 when $n_e$ becomes larger. 
Thus, a large range of temperatures, down to T$\sim$1000~K, is required to explain most of the observed ratios. Still, ratios as high as 1.0 cannot be explained by the Case B models. In addition, as pointed out in Edwards et al. (2013), the necessity to  maintain low line optical depths in the Case B models puts some restrictions to the applicability of this model to regions with high densities such as accretion funnels in YSOs.

\subsubsection{Kwan \& Fischer models}
\label{KF_section}
Kwan \& Fischer (2011) predicted hydrogen line ratios using a different approach than the Case B, in that they consider individual line optical depths. The inputs of these calculations are  the gas temperature, the hydrogen density, the ionization rate (which was not specified in the case B models) and the velocity gradient with respect to the emission length scale (see Kwan \& Fischer 2011 and Edwards et al. 2013 for more details). These parameters are chosen to replicate conditions in the region of winds and accretion flows in accreting YSOs. 

\begin{figure}[h]
\centering
\includegraphics[angle=-90,scale=0.3]{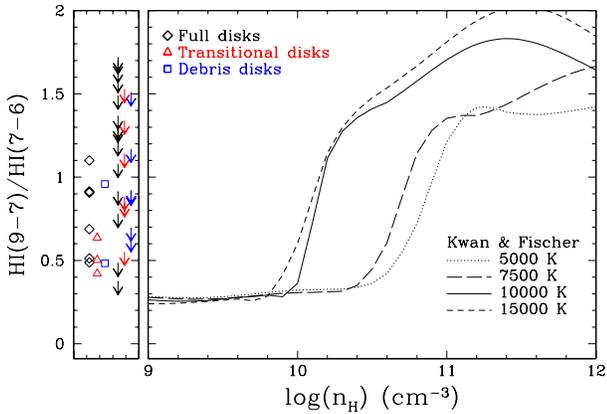}	
\caption{{\small HI(9-7)/HI(7-6) line ratios versus the hydrogen column density for the observed ratios. Kwan \& Fischer models for different temperatures as labeled. Symbols as in Fig.~\ref{caseBmodels}
\vskip 0.8cm
}
\label{KFmodels}}
\end{figure}

We show in Figure~\ref{KFmodels} the comparison between the KF model predictions for the {\hi}(9-7)/{\hi}(7-6) line ratio and the observations\footnote{Hydrogen line ratios for the Kwan \& Fischer models are publicly available: http://iopscience.iop.org/0004-637X/778/2/148/suppdata/data.tar.gz}. 
The input gas temperatures span from 5000~K to 15,000~K, and given that the (n$_e$)/(n$_H$) ratio is between 0.1 and 1.0, the hydrogen densities (n$_H$) cover similar electron densities as those shown in Fig.~\ref{caseBmodels} for the Case B models. 
The ratios suggested by the KF models are almost constant for all the temperatures until n$_H \sim 10^{10}$~cm$^{-3}$. As the density increases, the line optical depth increases as well, first for the (7-6) transition, which explains the steep increase in line ratio. For (n$_H$)$>$10$^{11}$cm$^{-3}$ both lines have large optical depths which explains the plateau.  
While the general behavior is similar for all temperatures, the specific transition density is $\sim10^{10}$~cm$^{-3}$ for T$\gtrsim$10,000~K and $\sim5\times10^{10}$cm$^{-3}$ for lower temperatures.  
Thus, unlike the Case B models, the KF models can explain all the observed line ratios with a limited range of temperatures (5000 -- 15,000~K) and relatively narrow range of gas densities (n$_H \sim 10^{10} - 10^{11}$cm$^{-3}$). 

\subsubsection{Comparison between model predictions and gas physical properties}
\label{sect_comp}

According to Martin (1996), who modeled the temperature and density profiles for gas channeled in magnetospheric accretion funnels, the typical temperatures in these regions are $\sim$6000~K, and hydrogen densities are as high as 10$^{9}$--10$^{10}$cm$^{-3}$. 
Muzerolle et al. (2001), considering the H$\alpha$ and H$\beta$ lines, found that the temperature of magnetospherically accreting gas ranges between 6000~K -- 12000~K  for a range of accretion rates between $10^{-6}$ -- $10^{-10}$ \Msun/yr. 
Recently, Edwards et al. (2013) examined near-IR hydrogen line ratios for a sample of T~Tauri stars. They find that the hydrogen density ($n_{H}$) in the line formation region is constrained within  2$\times$10$^{10}$--2$\times$10$^{11}$cm$^{-3}$, while the temperature is not well delineated. 

If the hydrogen mid-IR lines are tracing accretion at least in full and transitional disks, the observed {\hi}(9-7)/{\hi}(7-6) line ratios cannot be explained by the gas properties predicted by the case B models. In fact, as we have seen is Sect.~\ref{caseBmod} and Fig.~\ref{caseBmodels} the observed line ratios require gas colder than 5000~K irrespective of the electron densities. 
Such physical properties, and in particular such low temperatures, are unlikely in the magnetospheric accretion columns, in which we expect temperature of at least a factor of two higher and a much narrower range in density (e.g., Muzerolle et al. 2001). 

\begin{figure}[h]
\centering
\includegraphics[angle=-90,scale=0.3]{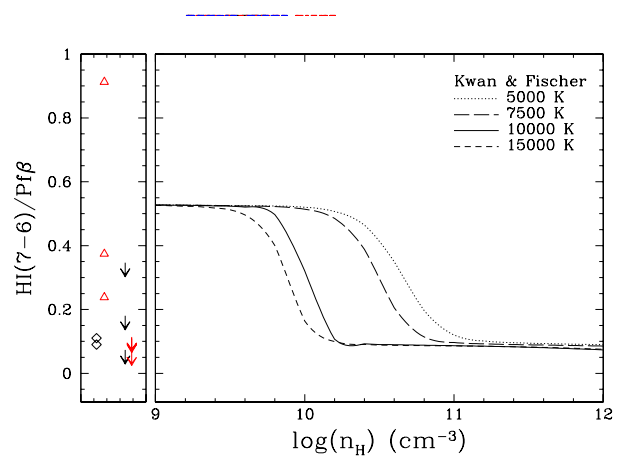}	
\caption{{\small HI(7-6)/Pf$\beta$ line ratios versus the hydrogen column density. Kwan \& Fischer models for different temperatures as labeled. Same symbols as Fig.~\ref{caseBmodels}. 
\vskip 0.8cm
}
\label{KFmodels_pfb}}
\end{figure}

On the other hand, the KF models suggest gas properties that better match the scenario in which the mid-IR hydrogen lines are produced in the accretion funnels because they predict an hydrogen column density ($n_{H}$) between $10^{10} - 10^{11}$ cm$^{-3}$, and a temperature range between 5,000 and 15,000~K (see Fig.~\ref{KFmodels}).
The analysis of the {\hi}(7-6)/Pf$\beta$ line ratios for the subsample of objects where these two lines have been observed adds additional constraints. In all but one object the observed line ratio is consistent with the KF model predictions for $n_{H}$ between $\sim10^{10} - \sim10^{11}$ cm$^{-3}$ and T$>$5,000~K (see fig.~\ref{KFmodels_pfb}). These results not only suggest that in most cases {\hi}(7-6) and Pf$\beta$ might have a common origin, but that this origin is from hot gas with relatively high hydrogen column density (as expected in the accretion funnels).  
The objects where both {\hi}(7-6) and Pf$\beta$ are detected, but the {\hi}(7-6)/Pf$\beta$ line ratio does not fit the Kwan \& Fischer models are VW~Cha and  TW~Hya, where the {\hi}(7-6)/Pf$\beta$ line ratios is $\sim$0.9.

We note that similarly stringent constraints on the gas physical properties cannot be placed when using shorter wavelengths hydrogen lines. In particular, we have analyzed model predictions (both case B and KF models) for the {\hi}(7-6)/H$\alpha$ and {\hi}(7-6)/Pa$\beta$ lines. We find that the observed ratios span a wider range of values, making the comparison with models inconclusive. This could be due to the different optical depths of the compared lines, which could affect the H$\alpha$ line as well as other short wavelength transitions, or to time variability, which may be more pronounced at shorter wavelengths. 
In the case of the {\hi}(7-6)/H$\alpha$ line ratio we must also take into account that Balmer series lines in general, and H$\alpha$ line in particular, have quite complicated profiles, with a variety of structures that include blueshifted absorption and line asymmetries (e.g., Basri \& Batalha 1990). 

\subsection{Chromosphere and High-Density Wind}
\label{sect:chrom}
The analysis carried out so far has shown that the {\hi} mid-IR lines are very unlikely disk tracers. Moreover, the correlation between the {\hi}(7-6) line and the accretion luminosity, and the {\hi} line ratios support an origin of these lines in the accretion funnels for full and transitional disks. 
However, Hollenbach \& Gorti (2009) discussed two other possibilities for the origin of {\hi} mid-IR lines: an internal wind shock with high speed ($v_{s} \gtrsim$ 100~km/s) and close to the wind origin ($\lesssim$1~AU), or the stellar chromosphere (see also Pascucci et al. 2007).   

An internal shock origin, even if plausible for full disks, would be very difficult to reconcile with the detection of {\hi} in the spectra of transitional and debris disks. 
Transition disks rarely show signs of outflows activity (see for example the optical forbidden line profiles in Hartigan et al. 1995), and debris disks have already dispersed most of their gas disk mass (e.g. Pascucci et al. 2006). 
While it is possible that mid-IR hydrogen lines trace different environments depending on the stellar/disk evolutionary stage, as shown for the {\neii} line (see Sect.~\ref{intro}), a wind shock origin cannot explain the detections in transitional and debris disks. 

\begin{figure}[h]
\centering
\includegraphics[scale=0.4]{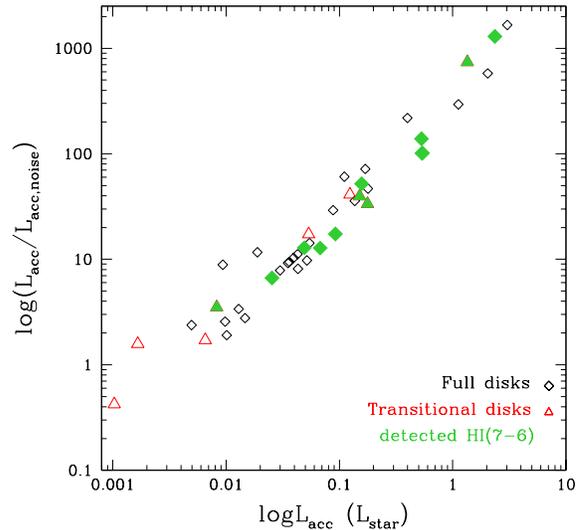}
\caption{{\small Contribution to the accretion luminosity from chromospheric emission (\Lacc/L$_{acc,noise}$), measured using the relation found by Manara et al. (2013), versus \Lacc\ for all the objects with T$_{{\rm eff}} <$4100~K. Symbols are as labeled. Filled symbols refer to objects where {\hi}(7-6) is detected. 
}
\label{chrom_test}}
\end{figure}

The mean chromospheric electron density in T Tauri stars is $n_e \sim 10^{11}$cm$^{-3}$ (Giampapa et al. 1981, or $10^{11} \lesssim n_H \lesssim 10^{12}$cm$^{-3}$ if  0.1$<$(n$_e$)/(n$_H$)$<$1, see Sect.~\ref{KF_section}).  According to the KF model at these high densities the {\hi}(7-6) and (9-7) transitions would be both optically thick, hence line ratios should be $\gtrsim$1. Only few sources in our sample sport such large line ratios.
Another way that we have employed to investigate the chromospheric origin of the {\hi} lines is to consider the effect of the chromospheric emission on the measured accretion luminosity. 
Manara et al. (2013), quantified this effect by analyzing a sample of high dispersion spectra of non-accreting low-mass T Tauri stars. Measuring the \Ha\ emission in non-accreting stars, and converting this value into an accretion luminosity, they find that the chromospheric activity can significantly contribute to the measured accretion luminosity when the latter is smaller that $10^{-3}$\Lsun, and that this contribution is strongly dependent on the object's effective temperature (hence spectral type). They derive a correlation between T$_{{\rm eff}}$ and L$_{acc,noise}$ (the latter representing the chromospheric contribution to the accretion luminosity). Unfortunately, their sample is limited to objects with spectral type in the range K7-M9.5, and none of our debris disks falls in this range. In the following we will focus on the 42 objects with T$_{{\rm eff}} <$4100~K and with an estimate of \Lacc\  (see Table~\ref{tab_all_prop}). 
In Fig.~\ref{chrom_test} we plot L$_{acc,noise}$ obtained using the correlation provided by Manara et al. (2013, their Eq. 2) versus  \Lacc\ as reported in  Table~\ref{tab_all_prop}. 13 of these objects also shown a detection in the {\hi}(7-6) line (green-filled symbols). 
There is a small number of targets, where the ratio is not more than 5, for which the chromospheric contribution could substantially influence the estimate of the accretion luminosity. However, in the majority of objects \Lacc\ is at least 5 times bigger than its chromospheric contribution, meaning that the latter is well within the estimated uncertainty on \Lacc. Moreover, the objects where the {\hi}(7-6) line is detected (identified as green-filled symbols in Fig.~\ref{chrom_test}) span all over the range of \Lacc.  

We can only test the chromospheric contribution over the accretion luminosity for $\sim$40\% of the total sample, where T$_{{\rm eff}} <$4100~K. For this subsample the previous analysis suggests that the {\hi} lines are not contaminated by the chromospheric contribution at a significant level. 
We could not extend this analysis to objects with higher temperatures (earlier spectral types) and to debris disks. 
{\it It is possible that the {\hi} mid-IR lines are tracing accretion in less evolved objects, while their contribution is only chromospheric in later evolutionary stages. }

The way to discriminate between accretion funnels, chromosphere and high-density wind is through ground-based high-resolution (10-20~km/s) spectroscopy. 
In fact, for {\hi} lines originating in accreting gas we will observe broad lines (FWHM$\sim$100~km/s, Muzerolle et al. 2001) close to the stellar velocity. A chromospheric component will be narrower ($\sim$40~km/s) and centered at the stellar velocity (e.g. Herczeg et al. 2007). An internal shock origin will instead produce {\hi} profiles that are as broad as 100~km/s and can be also blueshifted by the same amount. 

\subsection{A closer look at the Debris Disks} 
\label{section_DD}
\begin{figure}
\centering
\includegraphics[scale=0.42]{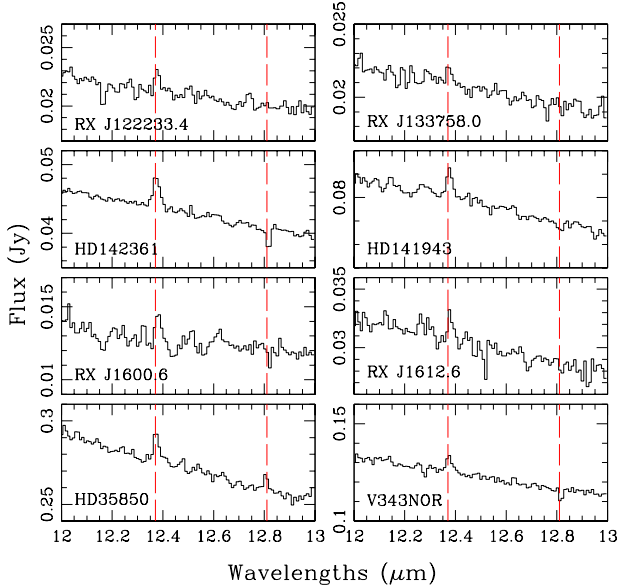}
\caption{{\small {\it Spitzer} spectra of the 8 Debris Disks where the {\hi}(7-6) line is detected. 
}
\label{spitz_dd}}
\end{figure}

\begin{figure}[h]
\centering
\includegraphics[scale=0.4]{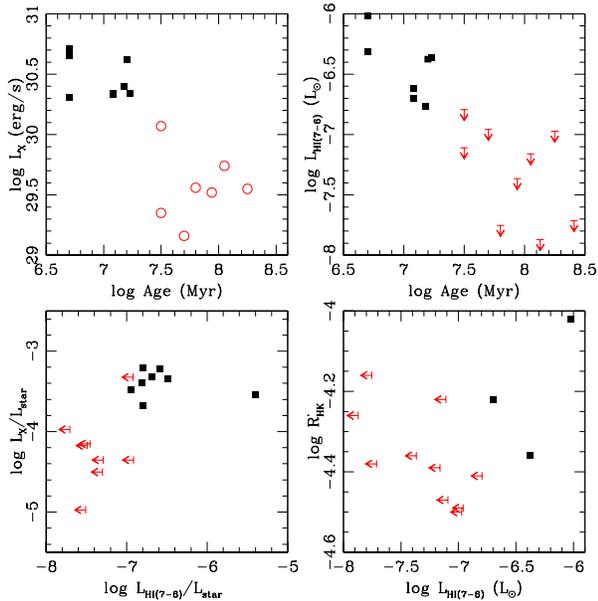}
\caption{{\small Top-left panel: X-ray luminosity versus Age. Black squares and red-open circles refer to objects with detected and not-detected {\hi}(7-6) line, respectively. Top-right panel: L$_X$/L$_{star}$ versus L$_{HI(7-6)}$/L$_{star}$. Bottom-left panel: chromospheric activity diagnostic R$_{HK}^{`}$ versus the {\hi} line luminosity. Bottom-right panel:  {\hi}(7-6) line luminosity versus Age.
}
\label{dd_prop_fig}}
\end{figure}

As already mentioned in the previous sections, this paper reports the first detection of {\hi}(7-6) and {\hi}(9-7) gas lines in debris disks (we also detect the {\neii} gas line in the {\it Spitzer} spectrum of HD~35850, as will be discussed in Rigliaco et al., in preparation).   
We show in Fig.~\ref{spitz_dd}  a compilation of the 8 debris disks line profiles where the {\hi}(7-6) line has been detected. 
In the following we will investigate the properties of all the debris disks we have re-reduced, either with detected {\hi}(7-6) lines or not (as reported in Tables~~\ref{tab_all_prop} and~\ref{flux_obj}). All these debris disks are observed within the FEPS\footnote{Formation and Evolution of Planetary Systems} {\it Spitzer} Survey (Meyer et al. 2006), and we collect the quantities plotted in Figure~\ref{dd_prop_fig} and reported in Table~\ref{dd_prop} from the FEPS Database\footnote{http://feps.as.arizona.edu/index.html}.

\begin{deluxetable}{l c c c c}
\tablecaption{Debris Disks Properties  \label{dd_prop}}
\tablehead{\colhead{Name} & \colhead{Age} & \colhead{log L$_X$}  & \colhead{R$^{`}_{HK}$} & Ref \\
	& (Myr) 	& (erg/s)	& & } 
\tabletypesize{\footnotesize}
\startdata
1RXS~J121236.4-552037 	&	17     &  	30.32	& \nodata &	 \ref{Mam02},\ref{FEPS}\\
1RXS~J122233.4-533347 	&	17    &  	30.34	& \nodata &	 \ref{Mam02},\ref{FEPS} \\
1RXS~J130153.7-530446 	&	17    &  	30.01	& \nodata &	  \ref{Mam02},\ref{FEPS}\\
HD~119269	&	17    &  	30.60	&		\nodata &	 \ref{Mam02},\ref{FEPS}\\
1RXS~J133758.0-413448 	&	15     &  	30.40	& \nodata &  \ref{Mam02},\ref{FEPS}	\\
1RXS~J161458.0-275013 	&	5     &  	30.05	& \nodata &	  \ref{Pre02},\ref{FEPS}\\
HD~142361 	&	 5    &  	30.71	&	-4.02 	&	\ref{Pre02},\ref{FEPS},\ref{Whi07}\\
RXJ1600.6-2159 	&	5     &  	30.31	&		\nodata & \ref{Pre02},\ref{FEPS}	\\
RXJ1612.6-1859 	&	5     &  	30.65	&		\nodata &	 \ref{Pre02},\ref{FEPS} 	\\
HD~35850 	&	 12    &  	30.34	&	-4.22 &	\ref{Zuc01},\ref{FEPS},\ref{Wri04} 	\\
V343Nor 	&	12     &  	30.33	&		\nodata &	 \ref{Zuc01},\ref{FEPS}	\\
HD~19668 	&	257     &  	\nodata	&	-4.38 &	\ref{FEPS},\ref{Wri04} \\
HD~25457 	&	112     &  	29.74	&	-4.39 &	\ref{Luh05},\ref{FEPS},\ref{Wri04}  \\
HD~37484 	&	50    &  	29.16	&	-4.49 &	\ref{FEPS},\ref{FEPS},\ref{Whi07}\\
HD~12039	& 	63	&	29.56	&	-4.16  &	\ref{FEPS},\ref{FEPS},\ref{Whi07}\\
HD~202917 	&	 32    &  	30.07	&	-4.22 &	\ref{FEPS},\ref{FEPS},\ref{Wri04}\\
HD~141943 	&	 16    &  	30.62	&	-4.36 &	\ref{FEPS},\ref{FEPS},\ref{Hen96}\\
HD~377 	&	87     &  	29.52	&	-4.36	 &	\ref{FEPS},\ref{FEPS},\ref{Wri04} \\
HD~17925	& 	55	&	29.07	&	-4.31 &	\ref{FEPS},\ref{FEPS},\ref{Bal96} \\
HD~32297	&	\nodata	& 	\nodata	&	\nodata &	\\
HD~61005 	&	135     &  		\nodata	&	-4.26 &	\ref{FEPS},\ref{Hen96} \\
HD~69830	&	1860	&	\nodata	&	-4.95 & \ref{Wri04},\ref{Wri04}	\\
HD~134319	&	186	&	\nodata	&	-4.35 & \ref{Wri04},\ref{Wri04}	\\
HD~216803	&	400	&	28.33	&	-4.27 &	\ref{FEPS},\ref{FEPS},\ref{Hen96}\\
AO~Men		&	12	&	30.16	& 	-3.75 & \ref{Zuc01},\ref{FEPS},\ref{Gre06}\\
  \enddata
\tablerefs{
\usecounter{minirefcount}
\mr{Baliunas et al. 1996;}{Bal96}
\mr{FEPS Database;}{FEPS}
\mr{Grey et al. 2006;}{Gre06}
\mr{Henry et al. 1996;}{Hen96}
\mr{Luhman et al. 2005;}{Luh05}
\mr{Mamajek et al. 2002;}{Mam02}
\mr{Preibisch et al. 2002;}{Pre02}
\mr{White et al. 2007;}{Whi07}
\mr{Wright et al. 2004;}{Wri04}
\mr{Zuckerman et al. 2001;}{Zuc01}
}
\end{deluxetable}

We first investigate a chromospheric nature for the mid-IR hydrogen lines. 
As shown in Figure~\ref{dd_prop_fig} (top-left panel), all the debris disks where the {\hi}(7-6) line is detected are younger and with higher X-ray luminosity.  The latter is a chromospheric indicator of magnetic activity (e.g. Feigelson et al. 2007), and its decay with age could let us conclude that the {\hi} line in debris disks are chromospherical, and as the magnetic activity decreases (with age) they are not detected anymore in older disks. 
Moreover, the X-ray emission of the debris disks with detected {\hi}(7-6) lines almost reaches the saturation level for chromospheric emission (L$_X$/L$_{star} \sim 10^{-3}$, e.g., Patten \& Simon 1996, Figure~\ref{dd_prop_fig}, top-right panel). This saturation level indicates that the chromospheric radiation emitted from the star cannot exceed a given fraction of the total stellar flux, and this limit is almost reached in all the observed debris disks, favoring again a chromospheric origin of the {\hi} lines.  
Another test to check whether or not {\hi} lines in debris disks are from chromospheric activity is shown in the bottom-left panel of Fig.~\ref{dd_prop_fig}, where we plot the stellar chromospherical activity indicator $R_{HK}^{'}$ (collected from White et al. 2007). This indicator measures the core emission in the Ca~{\sc ii} H and K lines, and the larger is the value, the more chromospherically active is the star (Noyes et al. 1984){\footnote {we note the reader that the $R_{HK}^{'}$ index may itself vary with time along the magnetic cycle of the star.}}. For the debris disks where this value has been measured there is no significant difference in terms of range covered by $R_{HK}^{'}$ as a function of {\hi} line luminosity, showing that stars with the same chromospheric activity (similar $R_{HK}^{'}$ value) may or may not have detected {\hi} line.  On the other hand, the object with strongest {\hi} line has the largest $R_{HK}^{'}$ value, which would support chromospheric origin in this case. However, the sample is limited to only three objects, and statistical studies are not possible. 

We also speculate on the possibility that, as for full and transitional disks, the detection of the mid-IR {\hi} lines in the {\it Spitzer} spectra of debris disks might trace accretion at very low levels. 
In fact, Fig.~\ref{dd_prop_fig}, bottom-right panel shows that the {\hi} line is detected in the youngest systems from our sample, for which low level of accretion may be still ongoing. 
While multiplicity studies are not complete for debris disks, it is intriguing to note that the two debris disks with a confirmed companion close enough to contribute to the observed {\it Spitzer} spectrum (see  Table~\ref{tab_all_prop} in Appendix A), have mid-IR hydrogen line detections. This raises yet another possibility that the {\hi}(7-6) line observed in debris disks is due to an accreting or chromospherically active companion, instead of the debris disk object itself. However, this remains a speculation at this point given the low number of detections.

In summary, even if a chromospheric origin appears the most likely based on the tests carried out so far, accretion cannot be ruled out. Only higher spectral resolution will be able to pin down the origin of mid-IR hydrogen lines.

\section{Summary and Outlook} 

We have re-reduced archival {\it Spitzer}/IRS spectra of 114 objects in different evolutionary stages, from full to transitional to debris disks, and we have investigated the origin of the {\hi} mid-IR lines. Our main findings are as follows. \\

1. After correcting for the disk water contamination, we detect {\hi}(7-6) line in 26 full disks, 12 transitional and 8 debris disks. The fainter {\hi}(9-7) line is detected in 6 full, 3 transitional and 2 debris disks.  \\

2. We find {\hi}(7-6) luminosities similar to the {\neii}  12.81\,\micron{} luminosities. These large luminosities exclude that the {\hi}(7-6) line traces the same hot disk surface probed via the {\neii} line. \\

3. Case B models cannot reproduce the observed {\hi}(9-7)/{\hi}(7-6) ratios except for T<3000K. On the other hand, KF models reproduce the observed ratios 
for reasonable assumption of temperatures and hydrogen column densities (n$_H$). Using the KF models the n$_H$ is fairly well constrained between $10^{10}<n_H<10^{11}$cm$^{-3}$ but the temperature is not. \\

4. For the subset of full and transitional disks we report a positive correlation between the {\hi}(7-6) line luminosity and the accretion luminosity. This trend, together with the results shown in point 3 suggest  an origin of these lines from gas accreting onto the star. \\

5. A pure chromospheric origin of the mid-IR {\hi} lines, can only be tested for sources with T$_{eff}<$4100~K (40\% of the total sample). For this subset of objects we find that the majority of them are not contaminated by chromospheric emission at a significant level.  \\

6. We report the first detection of {\hi} lines in {\it Spitzer} spectra of debris disks, which occurs only the in the youngest systems studied here (< 20 Myr). A chromospheric origin of these lines seems likely, but it is not possible to definitively rule out either a long-lasting accretion signature nature, or the presence of a non-resolved accreting companion.  \\

Ground-based high-resolution spectroscopy will enable testing this origin. 
Extending accretion indicators at mid-IR wavelengths may enable measuring mass accretion rates in less evolved, still embedded Class I objects (where the star is still surrounded by a remnant infall envelope and accretion disk). 
Still, care should be taken in converting the mid-IR {\hi} line luminosities into mass accretion rate, because the envelope surrounding Class I sources could absorb part of the emission. 
 More interestingly and on a speculative base, if the mid-IR {\hi} lines were confirmed as accretion indicators in debris disks, these lines may be used to extend mass accretion rate measurements to very low values. If the detected {\hi} lines are indeed tracing accretion, the luminosities we measure for the 8 debris disks correspond to mass accretion rates $\lesssim10^{-10}M_{\odot}$/yr. 
Optical and UV accretion diagnostics are not sensitive to such low mass accretion rates for sun-like stars (e.g., Manara et al. 2013). 

Finally, extending accretion indicators to the mid-IR will enable simultaneous observations of the {\neii} line (the only confirmed photoevaporative wind indicator) and the {\hi} line (an accretion indicator), making possible a direct comparison of the two most efficient disk dispersal mechanisms, photoevaporation and viscous accretion (Alexander et al. 2014).  \\

The authors thank Catherine Espaillat and Lynne Hillenbrand for providing the optical spectra and physical parameters of a few objects, and John Kwan for providing the mid-IR hydrogen line ratios. E.R. is supported by the NASA's Astrophysics Data Analysis Program research grant to I.P. (ID: NNX11AG60G).

\clearpage
\LongTables

\appendix
\section{Appendix}

\begin{figure}[h]
\vspace*{-5mm}
\plottwo{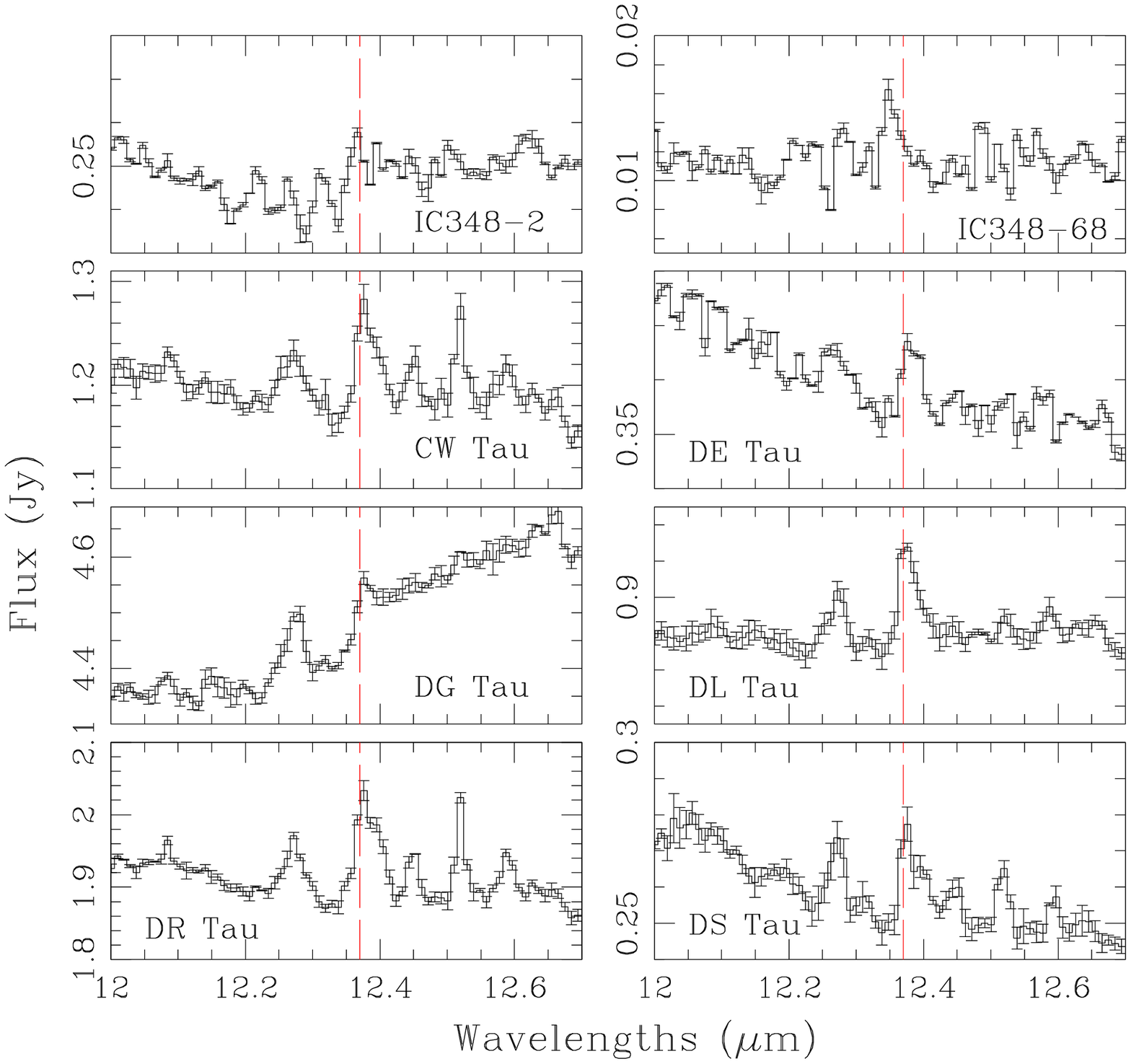}{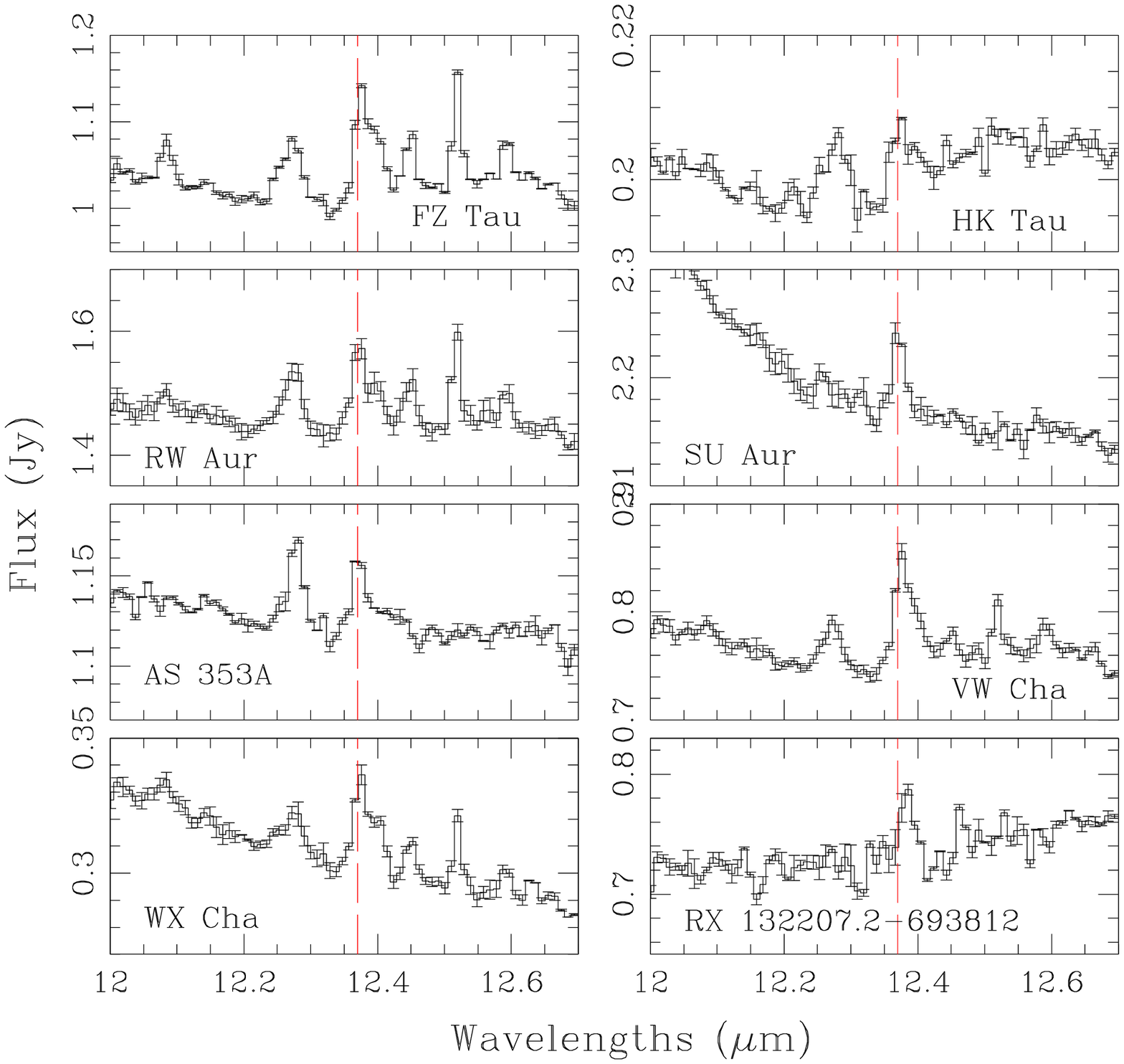}
\caption{\small IRS spectra around the HI(7-6) line for the objects where the line is detected
\label{all_HI_profile}}
\end{figure}

In Figure~\ref{all_HI_profile} we show the {\hi}(7-6) line for all the full and transitional disks where the line is detected. The {\hi}(7-6) line profile for debris disks are shown in Figure~\ref{spitz_dd}. 
In Table~\ref{tab_all_prop} we report the source properties, and Table~\ref{flux_obj} report the mid-IR line fluxes, errors and continuum levels around the three line of interest.

{\plottwo{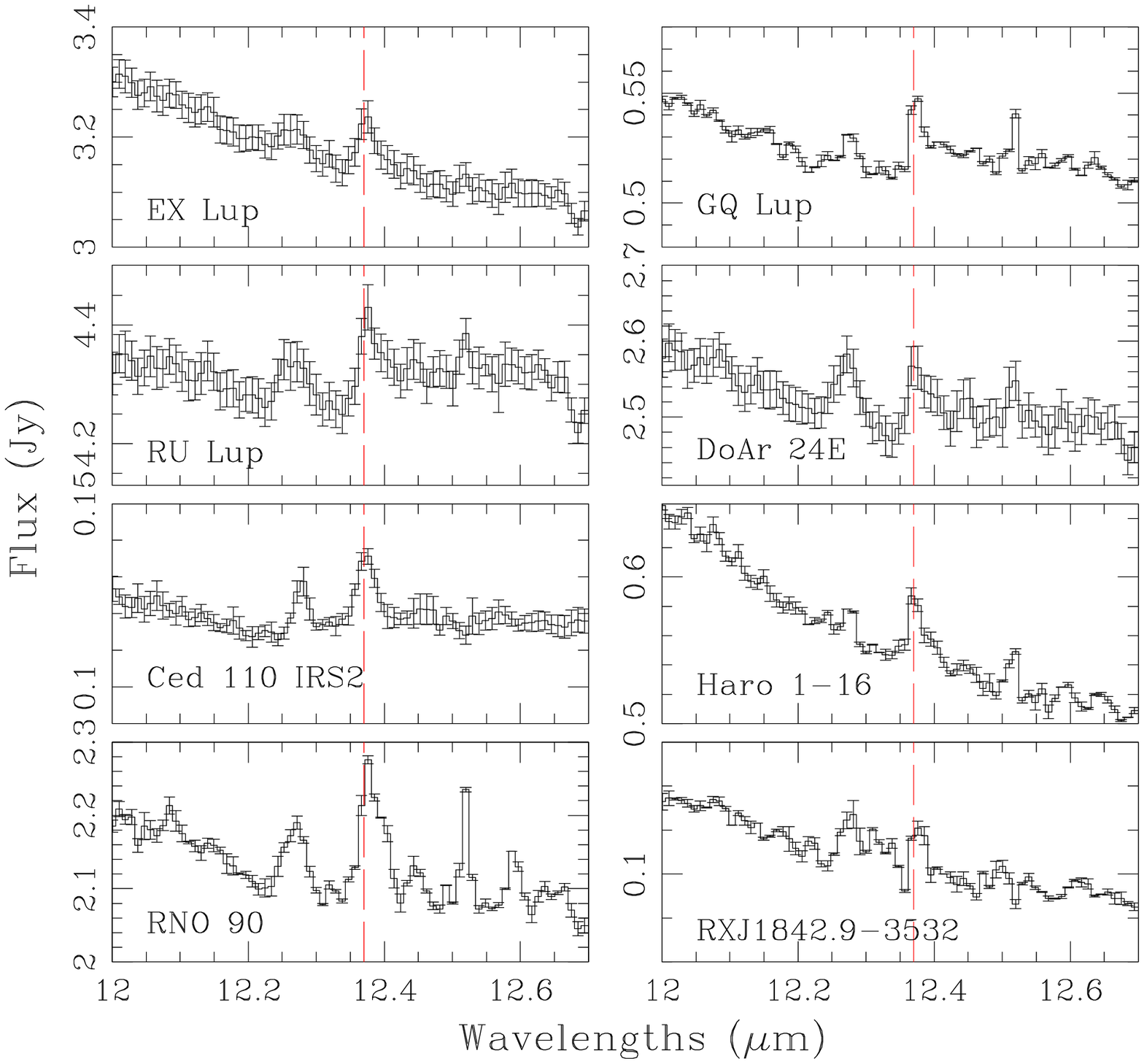}{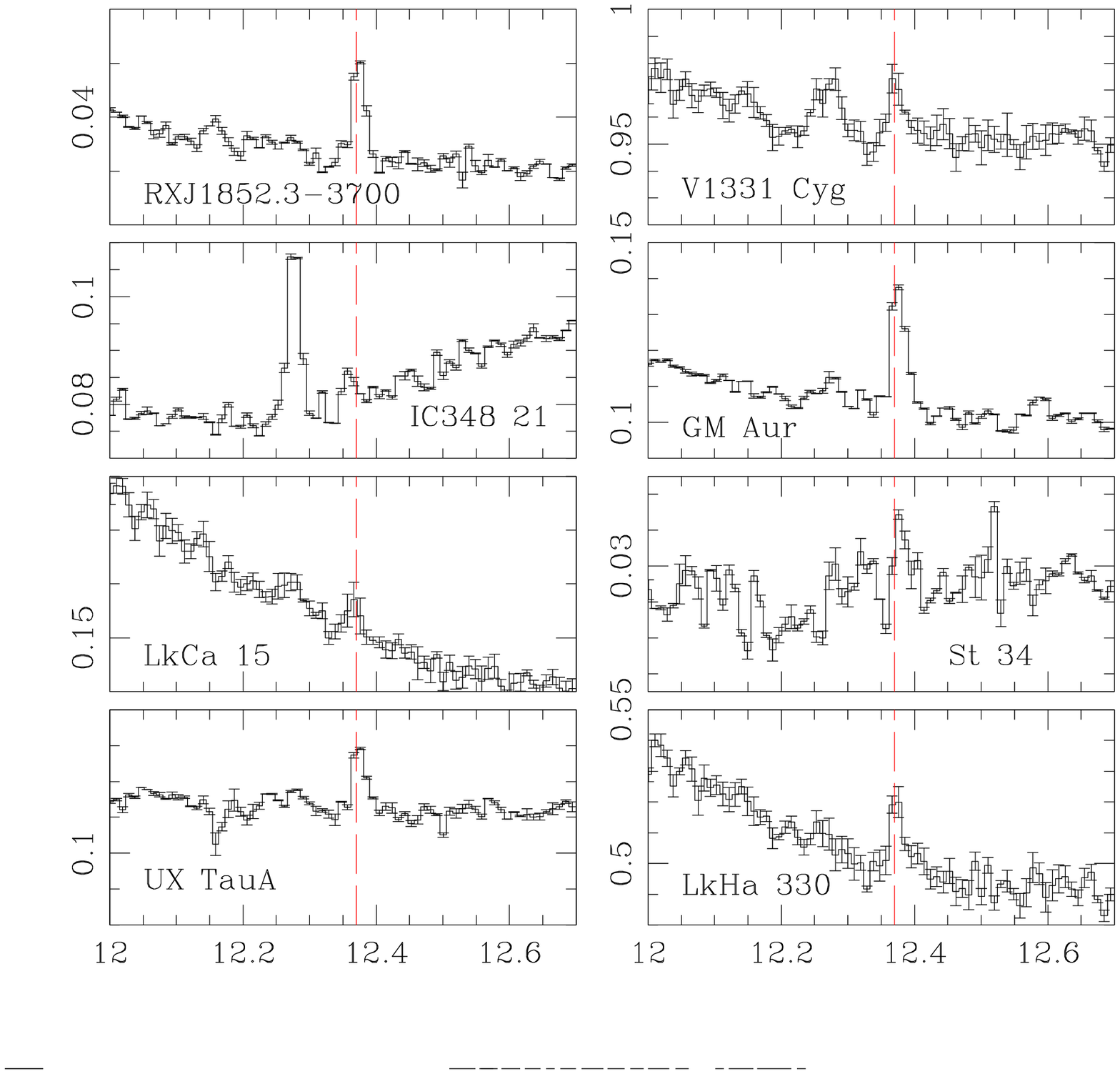}}\\
\centerline{\small Fig. A1. --- continued}
\epsscale{0.5}
\plotone{{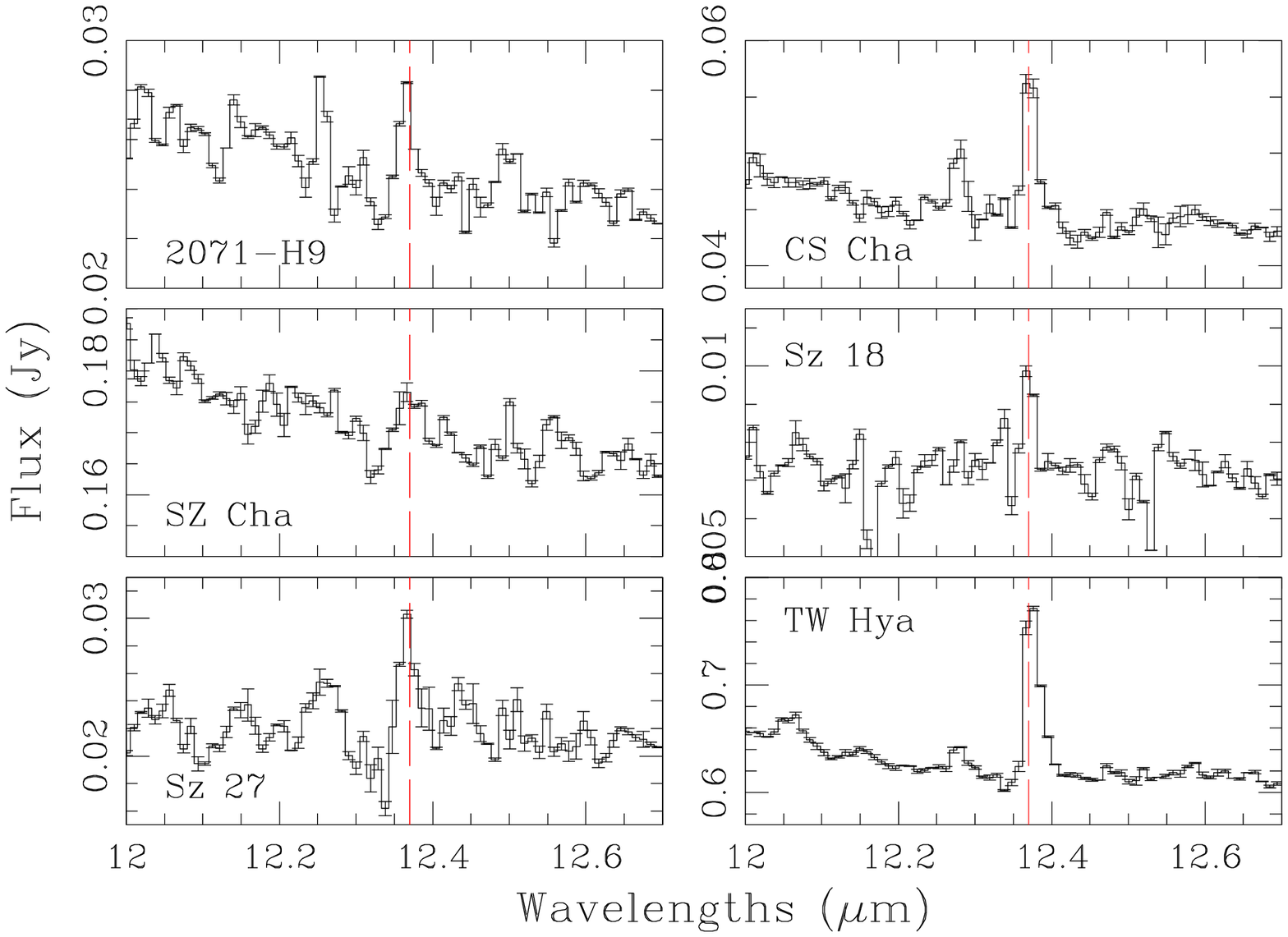}}\\
{\small Fig. A1. --- continued.}

\begin{deluxetable*}{l c c c c c c c c c c c c c c}
\tabletypesize{\footnotesize}
\tablewidth{0pt}
\tablecaption{Source Properties\label{tab_all_prop}}
\tablehead{\colhead{Name} 
	& \colhead{Association} 	
	&\colhead{Distance} 	
	& \colhead{SpTy} 	
	&\colhead{A$_V$}		
	&\colhead{T$_{\rm eff}$} 	
	& \colhead{M$_{star}$}	
	&\colhead{L$_{star}$} 
	& \colhead{R$_{star}$} 	
	& \colhead{H$\alpha$EW} 
	& \colhead{L$_{acc}$ } 
	& \colhead{Reference}
	\\
	& 	
	&\colhead{(pc)} 	
	&\colhead{}	
	& \colhead{} 
 	& (K) 
 	& (M$_{\odot}$) 
 	& (L$_{\odot}$) 
 	& (R$_{\odot}$) 
 	&  (\AA) 
 	& (L$_{\odot}$)  
 	& }
\startdata
\multicolumn{15}{c}{\emph{Full disks}} \\
\hline
IC348-2 &  IC 348 & 315 &      A2 &   3.8 &  8970 &   2.80 &  57.1 &   3.1 &  \nodata &   \nodata  & 
\ref{Her08},\ref{Esp12} 
\\	
IC348-6 & IC 348 &  315 &      G3 &   3.9 &  5830 &   2.40 &  16.6 &     4.0 &  \nodata &   \nodata &
\ref{Her08},\ref{Esp12}  
\\	
IC348-31 & IC 348 &   315 &      G6 &   8.6 &  5700 &   1.60 &     5.00 &   2.3 &  \nodata &   \nodata &
\ref{Her08},\ref{Esp12} 
\\
IC348-37 & IC 348 &   315 &      K6 &   2.8 &  4200 &   0.90 &   1.30 &   2.2 &  \nodata &   \nodata &
\ref{Her08},\ref{Esp12} 
\\
IC348-55 & IC 348 &  315 &     M0.5 &   8.5 &  3850 &   0.60 &     1.00 &   2.2 &  \nodata &   \nodata & 
\ref{Her08},\ref{Esp12} 
\\
IC348-68 &  IC 348 & 315 &      M3.5 &   2.1 &  3740 &   0.30 &   0.50 &     2.0 &  \nodata &   \nodata & 
\ref{Her08},\ref{Esp12} 
\\	
04216+2603 & Taurus &  140 &     M0 &     \nodata &  3500 & 0.28 &  0.02 & 0.3 &  \nodata &   \nodata &
\ref{Fur06},\ref{Ken98} 
\\
04385+2550 & Taurus &  140 &     M0 &   9.3 &  3850 & 0.57 & 3.67 & 4.0 &    20.0 & 0.50 & 
\ref{Fur06},\ref{Ken98} 
\\
AA~Tau$^*$ & Taurus &  140 &      K7 &  0.9 &  4060 & 0.73 &  1.38 & 2.3 &  37.1 & 0.07  
&
\ref{Fur06},\ref{Coh79}	
\\
BP~Tau &    Taurus &  140 &   K7 &  0.5 &  4060 & 0.73 &  1.62 & 2.4 &  40.1 & 0.07 
& 
\ref{Fur06},\ref{Coh79}	
\\ 
CW~Tau$^*$ & Taurus &  140 &   K3 &  2.2 &  4730 &  1.15 & 0.76 &   1.3 & 134.6 & 0.30 
& 
\ref{Fur06},\ref{Whi01}	
\\ 
CX~Tau &	Taurus &   140 &      M2.5 &  0.8 &  3490 & 0.36 &  0.62 & 2.0 &  19.8 & 0.003 
& 
\ref{Fur06},\ref{Coh79}	
\\
CY~Tau$^*$			&	Taurus	&	140 &     M1 &  0.9 &  3705 & 0.47 &  0.71 & 1.9 &  69.5 & 0.06  & 
\ref{Fur06},\ref{Coh79}	
\\
DD~Tau $^{m,*}$ 		&	Taurus	&	 140 &     M3 &  0.4 &  3415 & 0.29 & 0.13 & 1.0 & 181.8 & 0.05 &
\ref{Fur06},\ref{Whi01}, \ref{Coh79}	
\\
DE~Tau			&	Taurus	&	140 &    M1 &  1.6 &  3705 &  0.46 &  1.47 & 2.8 &    54.0 & 0.23  & 
\ref{Fur06},\ref{Whi01}, \ref{Coh79}	
\\
DG~Tau	&	Taurus	& 140 &     K6 &  1.4 &  4205 & 0.92 &  1.15 & 1.9 &    84.0 & 0.17  & 
\ref{Fur06},\ref{Whi01}	
\\
DK~Tau$^{m,*}$			&	Taurus	&	140 &     K7 &   1.4 &  4060 & 0.74 &  1.32 & 2.2 &  19.4 & 0.02 &
\ref{Fur06},\ref{Coh79},\ref{Str89}	
\\
DH~Tau$^*$ &	Taurus &   140 &     M0 &  1.0 &  3850 & 0.56 &  0.76 & 1.8 &  53.4 & 0.03
&
\ref{Fur06},\ref{Coh79} 
\\
DL~Tau$^*$			&	Taurus	&	140 &      K7 &     2.0 &  4060 & 0.75 &  1.16 & 2.2 &   105.0 & 0.62 & 
\ref{Fur06},\ref{Whi01}	
\\
DN~Tau$^*$			&	Taurus	&	140 &     M0 &  0.4 &  3850 & 0.55 &  1.62 & 2.7 &  11.9 & 0.016 & 
\ref{Fur06},\ref{Whi01}	
\\
DO~Tau$^*$			&	Taurus	&	140 &      M0 &  2.2 &  3850 &  0.56 &  1.05 & 2.2 &   100.0 & 0.19  & 
\ref{Fur06},\ref{Whi01}	
\\
DP~Tau$^{m,*}$			&	Taurus	&	140 &     M0.5 &  0.9 &  3850 & 0.56 &  0.85 & 1.9 &  85.4 & 0.05 & 
\ref{Fur06},\ref{Coh79}
\\
DQ~Tau$^{a,*}$			&	Taurus	&	140 &      M0 &  0.5 &  3850 & 0.56 &  0.93 & 2.0 & 112.9 & 0.04 & 
\ref{Fur06},\ref{Coh79}
\\
DR~Tau$^*$			&	Taurus	&	140 &     K5 &  0.5 &  4350 &  1.16 &  0.96 & 1.9 &   106.0 & 0.49  & 
\ref{Fur06},\ref{Whi01}	
\\
DS~Tau$^*$			&	Taurus	&	140 &     K5 &   0.9 &  4350 &   1.10 &  0.68 & 1.3 &  58.7 & 0.16 & 
\ref{Fur06},\ref{Whi01}	
\\
FM~Tau$^*$ &	Taurus&   140 &      M0 &   0.9 &  3850 &  0.57 &  0.52 & 1.5 &  70.9 & 0.02
& 
\ref{Fur06},\ref{Coh79} 
\\
FN~Tau			&	Taurus	&	140 &    M5 &   1.4 &  3125 & 0.22 &  0.79 & 2.8 &  24.7 & 0.007 & 
\ref{Fur06},\ref{Coh79}
\\
FT~Tau$^*$			&	Taurus	&	140 &     M3 &   3.8 &  3415 & 0.31 & 0.16 & 1.0 & 254.3 & 0.49 & 
\ref{Fur06},\ref{Coh79},\ref{Reb10} 
\\
FZ~Tau$^*$			&	Taurus	&	140 &    M0 &  2.7 &  3850 & 0.57 &   0.50 & 1.4 &   200.0 & 0.26 & 
\ref{Fur06},\ref{Whi01}	
\\
HK~Tau$^m$			&	Taurus	&	140 &    M0.5 &  3.0 &  3850 & 0.56 &  1.12 & 2.2 &  29.3 & 0.028 & 
\ref{Fur06},\ref{Coh79}
\\
HN~Tau$^{m,*}$			&	Taurus	&	 140 &     K5 &  1.2 &  4350 & 0.78 &  0.22 & 0.8 &   163.0 & 0.15 & 
\ref{Fur06},\ref{Whi01} 
\\
IP~Tau &	Taurus &   140 &     M0 &  0.5 &  3850 & 0.59 &  0.17 & 0.9 &    11.0 & 0.005 
& 
\ref{Fur06},\ref{Whi01} 
\\
RW~Aur$^{m,*}$			&	Taurus	&	140 &    K3 &  0.4 &  4730 & 1.48 &   1.60 & 1.8 &    76.0 & 0.924 & 
\ref{Fur06},\ref{Whi01} 
\\
SU~Aur			&	Taurus	&	140 &     G2 &  0.9 &  5860 & 1.85 &   8.90 &  2.7 &   3.5 & 0.14 & 
\ref{Fur06},\ref{Coh79} 
\\
UY~Aur$^{m}$ &	Taurus &   140 &     K7 &  1.1 &  3850 & 0.60 &  0.91 & 2.1 &  72.8 & 1.02 & 
\ref{Fur06},\ref{Coh79},\ref{Whi01} 
\\
AS~353A$^m$			&	Perseus	&	150 &     M3 &   2.1 &  3415 & 0.32 &  0.68 & 2.2 & 124.5 & 1.60 &
\ref{Pra03},\ref{Coh79}
\\
LkHa326 &	Perseus &   250 &      M0 &  3.1 &  3800 & 0.49 &  0.05 &  0.5 &  52.7 & 0.10 
&
\ref{Cas96},\ref{Coh79}
\\
2071-H13/ NGC2068		&	NGC2068 &	400 &     K1 &   2.9 &  5080 &  1.68 & 2.63 & 2.0 &  36.7 & 0.42 &  
\ref{Esp12}
\\	
2071-S1/ NGC2068 &  NGC2068 & 400 &     K4 &   4.1 &  4600 &   1.60 &   4.10 &   3.1 &   3.7 & 0.005 &  
\ref{Esp12}
\\	
RX~J1111.7-7620$^m$	&	Cha		&	160 &     K1 &   1.5 &  5080 & 1.78 & 3.09 & 2.2 &   7.6 & 0.04 &
\ref{Sil06},\ref{Wal92}
\\
SX~Cha$^{m,*}$			&	Cha		&	160 &      M0.5 &  1.1 &  3850 & 0.57 & 0.61 & 1.6 &  33.2 & 0.02 &
\ref{Fur11},\ref{Gue97}
\\
SZ50 &	Cha &   160 &     M3 &  2.1 &  3350 &   0.90 &  0.78 &     2.5 &    46.0 & 0.015 
&
\ref{Fur11},\ref{Hug92}
\\
TWCha  &	Cha&   160 &     M0 &   1.2 &  3850 & 0.56 &   3.10 & 3.7 &  28.3 & 0.04 
&
\ref{Fur11},\ref{Gue97}
\\
VW~Cha$^{m,*}$			&	Cha 		&	 160 &    K5 &  2.4 &  4350 & 1.14 &  4.02 & 3.3 &  71.7 & 0.81 & 
\ref{Fur11},\ref{Rei96} 
\\
VZ~Cha$^*$			&	Cha 		&	 160 &      K6 &  0.5 &  4205 & 0.93 & 0.84 &  1.6 &  71.4 & 0.099 &
\ref{Fur11},\ref{Rei96}
\\
WX~Cha$^{m,*}$			&	Cha 		&	 160 &    K7 &  2.1 &  4060 & 0.77 & 0.77 &  1.8 &  65.5 & 0.07 &
\ref{Fur11},\ref{App83}
\\
XX~Cha$^{m,*}$			&	Cha 		&	 160 &      M2 &  0.6 &  3560 & 0.37 & 0.18 & 1.1 & 133.5 & 0.03 &
\ref{Fur11},\ref{App83} 
\\
Hen-3-600A & TWHya &    56 &   M3 &     0.0 &  3415 & 0.26 &  0.06 & 0.6 &  21.8 & 0.007 
&
\ref{Uch04} 
\\
TWA10 & TWHya &   100 &    M2 &     0.0 &  5500 &   0.90 &   0.60 &     1.0 &  \nodata &   \nodata
&
\ref{Zuc04}
\\
1RXS~J132207.2-693812	&	LCC 	&	 86 &      K1 &  0.6 &  5035 &  1.23 &  1.14 &   1.4 &  19.5 & 0.07 & 
\ref{Sil06},\ref{Whi07}
\\ 
EX~Lup$^*$			&	Lupus	&	 190 &     M0 &     0.0 &  3850 & 0.55 & 1.31 & 2.4 &  31.3 & 0.06 & 
\ref{Lah07},\ref{Rei96} 
\\
GQ~Lup$^*$			&	Lupus	&	190 &      K7 &   0.4 &  4060 & 0.76 & 4.53 &  4.1 &  31.7 & 0.31 & 
\ref{Lah07},\ref{Rei96}
\\
HT~Lup$^m$ &	Lupus &   190 &    K3 &     0.0 &  4730 & 1.85 &     4.00 & 2.8 &   7.3 & 0.11
& 
\ref{Lah07},\ref{Rei96},\ref{Woi01}
\\
RU~Lup$^*$			&	Lupus	&	 190 &     K7 &   0.7 &  4060 & 0.74 & 1.28 & 2.2 &   136.0 & 0.69 & 
\ref{Lah07},\ref{Rei96}
\\
1RXS~J161411.0-230536$^m$	& Upper Scorpius	&  140 &      K0 &  1.1 &  5250 & 1.62 &  2.95 & 2.0 &  \nodata &   \nodata &
\ref{Sil06},\ref{Mam04} 
\\
HD~143006 &   Upper Scorpius  & 169 &    G7 &  0.6 &  5800 &  1.12 &   3.30 &     1.7 & 10.7 & 0.19 & 
\ref{Sil06},\ref{Mam04} 
\\
DoAr 24E$^{m,*}$			&	$\rho$Ophiucus	&	  160 &     G7 &   6.1 &  5630 & 2.15 & 10.75 & 3.3 &   7.1 & 0.24  &
\ref{Fur11},\ref{Wil05}  
\\f
DoAr 25$^*$			&	$\rho$Ophiucus	&	 160 &    K5 &   2.9 &  4350 & 1.11 & 2.19 & 2.3 &    12.0 & 0.05 &
\ref{Cie10},\ref{Wil05} 
\\
SR21$^{m}$ &	$\rho$Ophiucus &   125 &     G2.5 &     9.0 &  5950 & 2.67 &    28.0 & 4.7 &  \nodata &   \nodata 
&
\ref{Pra03}
\\
WaOph6$^*$			&	$\rho$Ophiucus	&	160 &    K6 &     1.0 &  4205 & 0.90 & 1.46 & 2.3 &    35.0 & 0.02  & 
\ref{Lah07},\ref{Wal86} 
\\
Ced110-IRS4 &	$\rho$Ophiucus&   160 &     K2 &    \nodata &  4800 & 0.85 &   0.40 & 0.9 &  \nodata &   \nodata  &
\ref{Pon05}
\\
Ced110-IRS2 &	$\rho$Ophiucus&   160 &     K3 &   3.3 &  4730 & 2.23 &  9.64 &  4.4 &  \nodata &   \nodata & 
\ref{Fur11} 
\\ 
Haro~1-16$^*$		&	$\rho$Ophiucus	&	160 &    K3 &  1.2 &  4730 & 1.80 &  3.39 &  2.7 &    59.0 & 0.24  &  
\ref{Fur06},\ref{Coh79}
\\
RNO~90$^*$			&	$\rho$Ophiucus	&	 160 &    G5 &     1.0 &  5770 & 1.28 & 2.74 & 1.6 &    76.0 & 0.47 & 
\ref{Lah07},\ref{Lev88} 
\\
EC~82			&	Serpens	& 	415 &     K7 &     1.0 &  4060 & 0.75 &  3.21 & 3.3 &     8.0 & 0.03 & 
\ref{Lah07},\ref{Coh79}
\\
RX~J1842.9-3532	&	CrA		& 	 136 &     K2 &  1.1 &  4900 &  1.33 & 1.29 & 1.5 &  27.2 & 0.06 &
\ref{Sil06},\ref{Whi07}
\\
RX~J1852.3-3700	&	CrA  		& 	136 &     K3 &  1.0 &  4730 & 1.27 &     1.00 & 1.5 &  27.4 & 0.06 &
\ref{Sil06},\ref{Whi07} 
\\	 
V1331Cyg & Cyn &   550 &     G8 &     0.0 &  5500 & 2.81 &    21.0 & 4.8 &  48.4 & 2.01
&
\ref{Coh79},\ref{Pet14} 
\\
\hline
\multicolumn{15}{c}{\emph{Transitional disks}} \\
\hline
IC348-21			&	IC 348	&	315 &     K0 &   4.7 &  5250 &   1.60 &   3.80 &   2.4 &   6.4 & 0.08  & 
\ref{Esp12}
\\	
IC348-67			&	IC 348	&	315 &    M0.75 &     2.0 &  3720 &   0.50 &   0.50 &   1.8 &  32.5 & 0.03 & 
\ref{Esp12} 
\\
IC348-72 &	IC 348&   315 &     M2.5 &     3.0 &  3580 &   0.40 &   0.70 &   2.1 &     2.0 & 0.0007  & 
\ref{Esp12} 
\\
IC348-133 &	IC 348&   315 &    M5 &   3.6 &  3200 &   0.20 &   0.20 &   1.5 &  \nodata &   \nodata 
& 
\ref{Esp12} 
\\
DM~Tau			&	Taurus	&	140 &     M1 &     0.0 &  3705 &  0.47 &  0.37 & 1.3 & 138.7 & 0.046 & 
\ref{Fur06},\ref{Ken98} 
\\
FP~Tau			&	Taurus	&	140 &     M5 &  0.3 &  3125 & 0.22 &  0.62 & 2.5 &  38.5 & 0.001 & 
\ref{Fur06},\ref{Ken98}
\\
GM~Aur			&	Taurus	&	140 &    K3 &  1.2 &  4730 & 1.27 &     1.00 & 1.5 &  96.5 & 0.58  & 
\ref{Fur06},\ref{Ken98}
\\
LkCa~15$^{m}$	&	Taurus	&	 140 &    K5 &  1.0 &  4300 & 1.06 &  0.74 & 1.5 &    13.0 & 0.027 & 
\ref{Fur06},\ref{Ken98}
\\
St~34$^m$			&	Taurus	&	140 &   M3 &  0.2 &  3415 & 0.24 & 0.01 & 0.3 &  51.6 & 0.016  & \ref{Lah07},\ref{Whi05}
\\
UX~Tau A$^m$		&	Taurus	&	140 &    K5 &  0.3 &  4350 & 1.17 &  0.91 & 1.9 &   9.5 & 0.03 & 
\ref{Fur06},\ref{Her88}
\\
V836Tau &	Taurus&   140 &     K7 &   0.9 &  4060 & 0.81 &  0.34 & 1.2 &  55.0 &   0.16 & 
\ref{Fur06},\ref{Ken98},\ref{Ngu09}
\\
LkHa~330			&	Perseus	&	 250 &    G3 &  1.8 &  5830 & 1.25 &  2.78 &  1.5 &  20.3 & 0.59 & 
\ref{Bro09},\ref{Fer95} 
\\
2071-H9/ NGC2068		&	NGC2068	&	400 &    K2 &   1.6 &  5250 & 1.41 & 2.03 & 1.7 &   1.2 & 0.004  &
\ref{Esp12}
\\
CHX 22 (T54)$^{m}$		&	Cha 		&	160 &     G8 &  1.2 &  5520 & 1.69 & 4.63 & 2.2 &  \nodata &   \nodata & 
\ref{Fur11},\ref{Wal92}, \ref{Ngu09}
\\
CS~Cha			&	Cha 		&	 160 &   K6 &   0.8 &  4205 &   0.90 &   1.50 &   2.3 &  54.3 & 0.39 & 
\ref{Fur11}, \ref{Rei96}
\\
SZ~Cha$^m$			&	Cha 		&	 160 &    K0 &   1.9 &  5250 &   1.40 &   1.90 &   1.7 &  \nodata &   \nodata &
\ref{Fur11}, \ref{Rei96}
\\
Sz~18 (T25)		&	Cha 		&	 160 &     M2.5 &   1.6 &  3560 & 0.38 & 0.32 & 1.4 &     8.0 & 0.003 & 
\ref{Fur11},\ref{Man11}
\\
Sz~27 (T35)		&	Cha 		&	160 &    K8 &   3.5 &  4060 & 0.80 & 0.47 & 1.2 &   100.0 & 0.08 & 
\ref{Fur11},\ref{Man11} 
\\
TW~Hya			&	TWHya	&	 56 &    M0 &     1.0 &  3850 & 0.57 & 0.64 & 1.7 &   194.0 & 0.09 & 
\ref{Cal02},\ref{Rei96}
\\
IM~Lup			&	Lupus	&	 190 &   M0 &   0.7 &  3850 & 0.56 & 3.52 & 3.9 &   4.7 & 0.02  & 
\ref{Pin08},\ref{Rei96}
\\
DoAr 21			&	$\rho$Ophiucus	&	 160 &     K1 &   6.6 &  5080 &   2.20 & 27.3 & 6.7 &   1.5 & 0.12 &
\ref{Muz10},\ref{Mer10}  
\\
\hline
\multicolumn{15}{c}{\emph{Debris disks}} \\
\hline
1RXS~J121236.4-552037 & LCC &  108 &    K0 &  0.2 &  5150 & 0.90 &   0.50 & 0.9 &  \nodata &   \nodata & 
\ref{Car03}, \ref{Che14}
\\
1RXS~J122233.4-533347	&	LCC	&	124 &    G0 &  0.3 &  5860 & 1.24 &  2.23 & 1.3 &  \nodata &   \nodata & 
\ref{Car03}, \ref{Che14}
\\
1RXS~J130153.7-530446 & LCC &   108 &     K2 & 0.2 &  4700 & 0.82 &   0.30 & 0.8 &  \nodata &   \nodata & 
\ref{Car03}, \ref{Che14}
\\
HD119269 & LCC &   85 &    G4 &  0.1 &  5700 &  1.07 &  1.31 &     1.2 &  \nodata &   \nodata 
& 
\ref{Car03}, \ref{Che14}
\\
1RXS~J133758.0-413448	&	UCL  & 98 &     K0 &  0.3 &  5250 &   0.90 & 0.68 & 0.9 &  \nodata &   \nodata & 
\ref{Car03}, \ref{Che14}
\\	
1RXS~J161458.4-27501 & Upper Scorpius &   114 &     G5 & 0.6 &  5250 &   0.90 &   0.70 &     1.0 &  \nodata &   \nodata 
&
\ref{Sil06},\ref{Whi07} 
\\
HD~142361$^m$		&	Upper Scorpius	&	 101 &   G3 &  0.5 &  5830 & 1.72 &  7.12 & 2.5 &  \nodata &   \nodata &
\ref{Car03}, \ref{Che14}
\\
RX~J1600.6-2159	 & Upper Scorpius	&	160 &    G9 &   0.8 &  5410 & 1.66 & 3.77 & 2.1 &  \nodata &   \nodata & 
\ref{Sil06},\ref{Whi07} 
\\
RX~J1612.6-1859$^m$	& Upper Scorpius	&	 127 &     K0 &  1.5 &  5250 & 1.50 &   2.40 & 1.8 &  \nodata &   \nodata  & 
\ref{Hil08},\ref{Whi07}
\\
HD~35850		&	$\beta$Pictoris  &	26.5 &     F7 &     0.0 &  6000 &   1.20 &  1.78 & 1.2 &  \nodata &   \nodata  &
\ref{Car03}, \ref{Che14}
\\
V343~Nor$^m$			&	$\beta$Pictoris &	38 &     K0 &     0.0 &  5250 & 0.90 & 0.67 & 0.9 &  \nodata &   \nodata  & 
\ref{Car03}
\\
HD19668 &  AB Dor  & 40 &     G8/K0 & 0.0 &  5500 &   0.90 &   0.60 &     0.8 &  \nodata &   \nodata & 
\ref{Car03}, \ref{Che14}
\\
HD25457 &  AB Dor &  19 &     F7 & 0.2 &  6000 &   1.20 &   2.50 &     1.5 &  \nodata &   \nodata & 
\ref{Car03}, \ref{Che14}
\\
HD37484 &  Columba &  60 &    F3 & 0.1 &  6740 &   1.30 &   3.90 &     1.5 &  \nodata &   \nodata & 
\ref{Car03}, \ref{Che14}
\\
HD12039$^{m}$ &  Tuc-Hor & 42 &     G4 & 0.2 &  6000 &     1.00 &     1.00 &     0.9 &  \nodata &   \nodata & 
\ref{Car03}, \ref{Che14}
\\
HD202917 & Tuc-Hor &   46 &    G7 & 0.1 &  5750 &   0.90 &   0.70 &     0.8 &  \nodata &   \nodata 
& 
\ref{Car03}, \ref{Che14}
\\
HD~141943		&	Field		&	 67 &     G0 &  0.1 &  6030 & 1.55 & 6.19 & 2.2 &  \nodata &   \nodata  & 
\ref{Car03}
\\
HD377 & Field & 39.6 &     G2 & 0.2 &  6250 &     1.00 &   1.50 &     1.0 &  \nodata &   \nodata & 
\ref{Car03}, \ref{Che14}
\\
HD17925 &  Field &  10 &     K1 &     0.0 &  5080 & 0.92 &  0.55 & 0.9 &  \nodata & \nodata & 
\ref{Car03}, \ref{Che14}		
\\
HD32297 &  Field & 112 &     A0 & 0.2 &  8000 &   1.90 &   6.20 &     1.3 &  \nodata &   \nodata & 
\ref{Car03}, \ref{Che14}
\\
HD61005 &  Field &  34 &    G8 &  0.1 &  5759 &   0.90 &   0.60 &     0.8 &  \nodata &   \nodata & 
\ref{Car03}, \ref{Che14}
\\
HD69830 &  Field &  13 &    G8 & 0.0 &  5500 &   0.90 &   0.60 &     0.85 &  \nodata &   \nodata & 
\ref{Car03}, \ref{Che14}
\\
HD134319 & Field &   44 &     G5 &     0.0 &  5770 &     1.00 &  0.72 &     0.8 &  \nodata &   \nodata & 
\ref{Car03}, \ref{Che14}
\\
HD216803 & Field &     8 &     K4 &     0.0 &  4600 &  0.66 &  0.19 &     0.7 &  \nodata &   \nodata & 
\ref{Car03}, \ref{Che14}
\\
AOMen &   Field & 39 &   K3 &     0.0 &  4435 &   0.70 &  0.26 &     0.9 &  \nodata &   \nodata & 
\ref{Car03}
\\

\enddata
\tablecomments{Objects marked with $^*$ are water-contaminated. $^a$=Spectroscopic Binary treated as single star (\ref{Kra11}). Stars marked with $^m$ are binary or multiple systems with the following separation: AS~353A 5.6$^{\prime\prime}$ (\ref{Kra11}), CHX~22 0.25$^{\prime\prime}$ (\ref{Laf08}), DD~Tau 0.56$^{\prime\prime}$ (\ref{Whi01}), DK~Tau 2.4$^{\prime\prime}$ (\ref{Kra11}), DoAr~24E 2.05$^{\prime\prime}$ (\ref{Rat05}), DP~Tau 0.11$^{\prime\prime}$ (\ref{Kra11}), HD~12039 0.15$^{\prime\prime}$ (\ref{Bil07}) - (this multiplicity has not been confirmed (\ref{Eva12})), HD~142361 0.73$^{\prime\prime}$ (\ref{Met09}) - (this multiplicity has not been confirmed (\ref{Met09}, \ref{Laf14})), HK~Tau 2.4$^{\prime\prime}$ (\ref{Lei93}), HN~Tau 3.1$^{\prime\prime}$ (\ref{Lei93}), HT~Lup 2.8$^{\prime\prime}$ (\ref{Woi01}), LkCa15 0.08$^{\prime\prime}$ (\ref{Kra12}), RX J1111.7-7620 0.27$^{\prime\prime}$ (\ref{Laf08}), RX J1612.6-1859 0.21$^{\prime\prime}$ (\ref{Met09}), 1RXS J161411.0-230536 0.22$^{\prime\prime}$ (\ref{Met09}), RW~Aur 1.4 $^{\prime\prime}$ (\ref{Kra11}), SR21 6.4$^{\prime\prime}$ (\ref{Rei93}), St~34 1.18$^{\prime\prime}$ (\ref{Whi14}), SX~Cha 2.2$^{\prime\prime}$ (\ref{Laf08}), SZ~Cha 5.12$^{\prime\prime}$ (\ref{Laf08}), UX~TauA 2.7,0.14$^{\prime\prime}$ (\ref{Duc99b}),  UY~Aur 0.89$^{\prime\prime}$ (\ref{Lei93}), V343NOR 10$^{\prime\prime}$ (\ref{Tor08}), VW~Cha 0.64,0.10 $^{\prime\prime}$ (\ref{Laf08}), WX~Cha 0.79 $^{\prime\prime}$ (\ref{Laf08}). 
}
\tablerefs{
\usecounter{minirefcount}
\mr{Appenzeller et al. 1983;}{App83}
\mr{Biller et al. 2007;}{Bil07}
\mr{Brown et al. 2009;}{Bro09}
\mr{Calvet et al. 2002;}{Cal02}
\mr{Carpenter \& Stauffer 2003}{Car03}
\mr{Casali \& Eiroa 1996}{Cas96}
\mr{Chen et al. 2014}{Che14}
\mr{Cieza et al. 2010;}{Cie10}
\mr{Cohen \& Kuhi 1979;}{Coh79}
\mr{Duchene et al. 1999;}{Duc99b}
\mr{Espaillat et al. 2012;}{Esp12}
\mr{Evans et al. 2012;}{Eva12}
\mr{Fernandez et al. 1995;}{Fer95}
\mr{Furlan et al. 2006;}{Fur06}
\mr{Furlan et al. 2011;}{Fur11}
\mr{Guenther \& Emerson 1997;}{Gue97}
\mr{Herbig \& Bell 1988;}{Her88}
\mr{Herbst, W. 2008;}{Her08}
\mr{Hillenbrand et al. 2008;}{Hil08}
\mr{Hughes \& Hartigan 1992}{Hug92}
\mr{Kenyon et al. 1998;}{Ken98}
\mr{Kraus et al. 2011;}{Kra11}
\mr{Kraus \& Ireland 2012;}{Kra12}
\mr{Lafreniere et al. 2008;}{Laf08}
\mr{Lafreniere et al. 2014;}{Laf14}
\mr{Lahuis et al. 2007;}{Lah07}
\mr{Leinert et al. 1993;}{Lei93}
\mr{Levreault 1988;}{Lev88}
\mr{Mamajek et al. 2004;}{Mam04}
\mr{Manoj et al. 2011}{Man11}
\mr{Merin et al. 2010;}{Mer10}
\mr{Metchev \& Hillenbrand 2009;}{Met09}
\mr{Muzerolle et al. 2010;}{Muz10}
\mr{Nguyen et al. 2012;}{Ngu09}	
\mr{Petrov et al. 2014}{Pet14}
\mr{Pinte et al. 2008;}{Pin08}
\mr{Pontoppidan \& Dullemond 2005;}{Pon05}
\mr{Prato et al. 2003;}{Pra03}
\mr{Ratzka et al. 2005;}{Rat05}
\mr{Rebull et al. 2010;}{Reb10}
\mr{Reipurth \& Zinnecker 1993}{Rei93}
\mr{Reipurth et al. 1996;}{Rei96}
\mr{Silverstone et al. 2006;}{Sil06}
\mr{Strom et al. 1989}{Str89}
\mr{Torres et al. 2008;}{Tor08}
\mr{Uchida et al. 2004}{Uch04}
\mr{Walter 1986;}{Wal86}
\mr{Walter 1992;}{Wal92}	
\mr{White \& Ghez 2001;}{Whi01}
\mr{White \& Hillenbrand 2005;}{Whi05}
\mr{White et al. 2007;}{Whi07}
\mr{White et al. in prep;}{Whi14}
\mr{Wilking et al. 2005;}{Wil05}
\mr{Woitas et al. 2001}{Woi01}
\mr{Zuckerman \& Song 2004}{Zuc04}
}
\end{deluxetable*}

\clearpage

\begin{deluxetable*}{l | c c c | c c c | c c c}
\centering
\tablecaption{Mid-IR line fluxes, errors and continuum level around the lines\label{flux_obj}} 
\tablehead{
    &  \multicolumn{3}{c}{{\neii}} &  \multicolumn{3}{c}{{\hi}(7-6)} &  \multicolumn{3}{c}{ {\hi}(9-7)} \\ 
\colhead{Name} & \colhead{Flux\tablenotemark{a}} 	&\colhead{1$\sigma$\tablenotemark{a}}		&\colhead{cont\tablenotemark{b}} 	& \colhead{Flux\tablenotemark{a}}	&\colhead{1$\sigma$\tablenotemark{a}}		&\colhead{cont\tablenotemark{b}} &\colhead{Flux\tablenotemark{a}} &\colhead{1$\sigma$\tablenotemark{a}}		&\colhead{cont\tablenotemark{b}} 
} \\
\startdata
\multicolumn{10}{c}{\emph{Full disks}} \\
\hline
IC348-2 	& 7.66E-15&  1.81E-15&     0.24&  4.05E-15&  7.45E-16&     0.25&  $<$6.46E-15&      \nodata&     0.33\\
IC348-6 &  $<$2.23E-15&      \nodata&    0.078&  $<$2.73E-15&      \nodata&    0.077&  $<$1.18E-14&      \nodata&     0.11\\
IC348-31&  $<$2.68E-15&      \nodata&    0.076&  $<$3.43E-15&      \nodata&    0.067&  $<$3.62E-15&      \nodata&    0.067\\
IC348-37&  $<$2.46E-15&      \nodata&    0.055&  $<$2.22E-15&      \nodata&    0.059&  $<$2.92E-15&      \nodata&    0.081\\
IC348-55 &  2.16E-15&  7.28E-16&    0.034&  $<$1.94E-15&      \nodata&    0.037&  $<$3.31E-15&      \nodata&    0.052\\
 IC348-68&  $<$1.41E-15&      \nodata&    0.011&  2.68E-15&  4.97E-16&    0.011&  $<$3.51E-15&      \nodata&    0.012\\
04216+2603&  $<$3.41E-15&      \nodata&     0.20&  $<$4.77E-15&      \nodata&     0.20&  $<$7.17E-15&      \nodata&     0.23\\
04385+2550$^*$ 	&  4.33E-15&  1.66E-15&     0.35&  $<$7.61E-15&      \nodata&     0.34&  $<$1.08E-14&      \nodata&     0.38\\
AA~Tau$^*$	& 1.32E-14&  2.14E-15&     0.31&  $<$7.61E-15&      \nodata&     0.32&  $<$1.63E-14&      \nodata&     0.41\\
BP~Tau$^*$	& $<$5.61E-15&      \nodata&     0.36&  $<$6.70E-15&      \nodata&     0.38&  $<$1.80E-14&      \nodata&     0.51\\
CW~Tau$^*$	& $<$1.71E-14&      \nodata&      1.13&  3.45E-14  (5.73E-14)&  4.20E-15&      1.18&  $<$3.73E-14&      \nodata&      1.33\\
CX~Tau&  $<$2.38E-15&      \nodata&     0.17&  $<$3.01E-15&      \nodata&     0.17&  $<$5.55E-15&      \nodata&     0.20\\
CY~Tau$^*$ 	& 2.22E-15&  9.68E-16&     0.15&  $<$4.93E-15&      \nodata&     0.16&  $<$3.53E-15&      \nodata&     0.18\\
DD~Tau$^*$  & 2.02E-14&  4.73E-15&     0.93&  $<$1.37E-14&      \nodata&     0.96&  $<$5.02E-14&      \nodata&      1.14\\
DE~Tau 	& $<$5.75E-15&      \nodata&     0.35&  7.81E-15&  1.40E-15&     0.35&  $<$1.34E-14&      \nodata&     0.49\\
DG~Tau 	& 2.43E-13&  1.13E-14&      4.64&  4.49E-14&  6.49E-15&      4.47&  $<$5.44E-14&      \nodata&      4.52\\
DK~Tau$^*$	& $<$9.78E-15&      \nodata&     0.78&  $<$2.46E-14&      \nodata&     0.88&  $<$1.54E-14&      \nodata&      1.24\\
DH~Tau$^*$  & $<$3.99E-15&      \nodata&     0.12&  $<$4.01E-15&      \nodata&     0.12&  $<$1.04E-14&      \nodata&     0.16\\
DL~Tau$^*$ 	&1.24E-14&  2.15E-15&     0.87&  7.12E-15 (3.24E-14)&  2.09E-15&     0.88&  1.58E-14 (1.84E-14)&  2.52E-15&     0.93\\
DN~Tau$^*$ 	& 3.83E-15&  5.38E-16&     0.28&  $<$3.22E-15&      \nodata&     0.29&  $<$2.92E-15&      \nodata&     0.31\\
DO~Tau$^*$	& $<$1.51E-14&      \nodata&      1.89&  $<$1.75E-14&      \nodata&      1.94&  $<$1.76E-14&      \nodata&      2.17\\
DP~Tau$^*$ 	&1.16E-13&  3.63E-15&     0.63&  $<$2.42E-14&      \nodata&     0.66&  $<$2.44E-14&      \nodata&     0.78\\
DQ~Tau$^*$	&9.06E-15&  3.33E-15&     0.60&  $<$1.14E-14&      \nodata&     0.61&  $<$1.19E-14&      \nodata&     0.64\\
DR~Tau$^*$ 	&  $<$2.12E-14&      \nodata&      1.88&  3.46E-14 (1.03E-13)&  6.93E-15&      1.88&  1.69E-14 (4.08E-14)&  8.65E-15&      2.10\\
DS~Tau$^*$	&$<$6.33E-15&      \nodata&     0.24&  4.23E-15  (1.28E-14)&  1.10E-15&     0.25&  $<$6.32E-15&      \nodata&     0.36\\
FM~Tau$^*$ &  $<$2.26E-15&      \nodata&     0.25&  $<$4.85E-15&      \nodata&     0.27&  $<$6.71E-15&      \nodata&     0.34\\
FN~Tau 	& 5.44E-15&  1.23E-15&     0.68&  $<$5.71E-15&      \nodata&     0.67&  $<$5.59E-15&      \nodata&     0.83\\
FT~Tau$^*$ 	&5.97E-15&  1.45E-15&     0.25&  $<$4.54E-15&      \nodata&     0.26&  $<$1.07E-14&      \nodata&     0.31\\
FZ~Tau$^*$ 	& $<$1.75E-14&      \nodata&     0.98&  2.46E-14 (8.65E-14)&  3.42E-15&      1.02&  2.23E-14 (3.17E-14)&  7.12E-15&      1.15\\
HK~Tau 	&5.51E-15&  1.08E-15&     0.20&  3.44E-15&  7.21E-16&     0.20&  $<$4.42E-15&      \nodata&     0.24\\
HN~Tau$^*$ 	&2.59E-14&  4.81E-15&     0.95&  $<$1.66E-14&      \nodata&      1.01&  $<$4.05E-14&      \nodata&      1.36\\
IP~Tau&  $<$6.22E-16&      \nodata&    0.013&  $<$8.16E-16&      \nodata&    0.014&  $<$1.94E-15&      \nodata&    0.020\\
RW~Aur$^*$	&$<$2.00E-14&      \nodata&      1.43&  1.96E-14 (7.16E-14) &  5.92E-15&      1.46&  $<$3.23E-14&      \nodata&      1.76\\
SU~Aur 	&1.78E-14&  3.63E-15&      2.14&  3.27E-14&  3.87E-15&      2.17&  $<$7.15E-14&      \nodata&      3.02\\
UYAur&  $<$2.49E-14&      \nodata&      2.36&  $<$3.39E-14&      \nodata&      2.38&  $<$8.22E-14&      \nodata&      2.67\\
AS~353A 		& 1.11E-14&  2.50E-15&      1.11&  1.96E-14&  3.07E-15&      1.13&  $<$1.80E-14&      \nodata&      1.26\\
LkHa326&  4.74E-15&  1.23E-15&     0.33&  $<$8.34E-15&      \nodata&     0.33&  $<$1.27E-14&      \nodata&     0.36\\
2071-H13/ NGC2068 & 1.01E-15&  2.43E-16&    0.024&  $<$9.83E-16&      \nodata&    0.026&  $<$1.71E-15&      \nodata&    0.038\\
NGC20682071-S1&  $<$1.97E-15&      \nodata&    0.046&  $<$1.77E-15&      \nodata&    0.050&  $<$2.23E-15&      \nodata&    0.061\\
RX~J1111.7-7620 &9.97E-15&  1.28E-15&     0.22&  $<$5.96E-15&      \nodata&     0.23&  $<$9.55E-15&      \nodata&     0.27\\
SX~Cha$^*$ 	&$<$7.08E-15&      \nodata&     0.49&  $<$1.34E-14&      \nodata&     0.51&  $<$1.42E-14&      \nodata&     0.65\\
SZ50&  $<$3.20E-15&      \nodata&     0.20&  $<$4.78E-15&      \nodata&     0.21&  $<$7.16E-15&      \nodata&     0.22\\
TWCha&  $<$4.35E-15&      \nodata&    0.062&  $<$4.01E-15&      \nodata&    0.055&  $<$6.63E-15&      \nodata&    0.042\\
VW~Cha$^*$ 	&  3.13E-14&  3.50E-15&     0.73&  3.74E-14 (5.98E-14)&  2.79E-15&     0.76&  1.91E-14 (2.64E-14)&  5.18E-15&     0.95\\
VZ~Cha$^*$ 	& $<$5.92E-15&      \nodata&     0.37&  $<$8.84E-15&      \nodata&     0.40&  $<$8.61E-15&      \nodata&     0.46\\
WX~Cha$^*$ 	&$<$5.08E-15&      \nodata&     0.28&  1.29E-14 (2.62E-14) &  1.34E-15&     0.30&  $<$1.01E-14&      \nodata&     0.43\\
XX~Cha$^*$ 	&4.57E-15&  7.48E-16&     0.13& $<$5.66E-15&      \nodata&     0.14&  $<$6.46E-15&      \nodata&     0.16\\
Hen-3-600A&  $<$6.62E-15&      \nodata&    0.099&  $<$4.78E-15&      \nodata&    0.095&  $<$1.08E-14&      \nodata&     0.11\\
TWA10&  $<$2.44E-15&      \nodata&    0.018&  $<$2.09E-15&      \nodata&    0.019&  $<$3.05E-15&      \nodata&    0.022\\
RX~132207.2-693812 &$<$2.59E-14&      \nodata&     0.75&  3.46E-14&  6.13E-15&     0.74&  3.17E-14&  1.23E-14&     0.84\\
EX~Lup$^*$	& $<$1.63E-14&      \nodata&      3.13&  1.71E-14 (5.34E-14) &  6.85E-15&      3.21&  $<$4.43E-14&      \nodata&      4.12\\
GQ~Lup$^*$ 	&  $<$6.38E-15&      \nodata&     0.49&  4.39E-15 (1.23E-14) &  2.29E-15&     0.52&  $<$9.91E-15&      \nodata&     0.65\\
HTLup&  $<$1.90E-14&      \nodata&      2.17&  $<$1.96E-14&      \nodata&      2.21&  $<$5.23E-14&      \nodata&      2.64\\
RU~Lup$^*$ 	& 4.54E-14&  8.41E-15&      4.29&  3.61E-14 (7.74E-14) &  8.24E-15&      4.36&  $<$6.13E-14&      \nodata&      4.80\\
RX~J161411.0-230536 &6.44E-15&  1.36E-15&     0.34&  $<$5.69E-15&      \nodata&     0.35&  $<$7.30E-15&      \nodata&     0.36\\
HD143006&  $<$1.08E-14&      \nodata&     0.69&  $<$1.63E-14&      \nodata&     0.74&  $<$3.68E-14&      \nodata&      1.12\\
DoAr~24E$^*$ 	&1.38E-14&  4.20E-15&      2.54&  1.83E-14 (3.96E-14) &  7.58E-15&      2.58&  $<$4.14E-14&      \nodata&      2.96\\
DoAr~25$^*$ 	& 2.87E-15&  9.27E-16&     0.21&  $<$7.78E-15&      \nodata&     0.22&  $<$3.84E-15&      \nodata&     0.24\\
SR21&  $<$2.84E-14&      \nodata&     0.79&  $<$3.63E-14&      \nodata&     0.75&  $<$6.01E-14&      \nodata&     0.91\\
Wa~Oph6$^*$ 	&8.55E-15&  1.83E-15&     0.82&  $<$2.25E-14&      \nodata&     0.82&  $<$1.65E-14&      \nodata&     0.92\\
Ced110IRS2&  $<$4.43E-15&      \nodata&     0.11&  1.92E-15&  6.24E-16&     0.12&  $<$6.97E-15&      \nodata&     0.15\\
Ced110-IRS4&  2.43E-14&  8.57E-16&    0.080&  $<$1.84E-15&      \nodata&    0.070&  $<$1.75E-15&      \nodata&    0.054\\
Haro~1-16$^*$ & $<$9.73E-15&      \nodata&     0.50&  1.05E-14  (2.16E-14) &  3.42E-15&     0.54&  $<$2.43E-14&      \nodata&      1.05\\
RNO~90$^*$ 	&2.10E-14&  1.08E-14&      2.05&  5.64E-14 (1.28E-13) &  6.22E-15&      2.10&  3.87E-14  (4.92E-14)&  9.83E-15&      2.49\\
EC~82 	& 1.39E-14&  3.05E-15&      1.32&  $<$1.48E-14&      \nodata&      1.45&  $<$3.93E-14&      \nodata&      2.57\\
RX~J1842.9-3532 &  4.75E-15&  7.21E-16&    0.092&  3.83E-15&  7.96E-16&     0.10&  $<$5.25E-15&      \nodata&     0.19\\
RX~J1852.3-3700 & 8.66E-15&  4.28E-16&    0.035&  4.62E-15&  3.26E-16&    0.037&  $<$1.74E-15&      \nodata&    0.055\\
V1331Cyg&  $<$3.30E-15&      \nodata&     1.01&  1.16E-14&  6.24E-16&     0.96&  $<$5.64E-15&      \nodata&     0.95\\
\hline
\multicolumn{10}{c}{\emph{Transitional disks}} \\
\hline
IC348-21 & 3.12E-15&  5.13E-16&    0.088&  3.71E-15&  9.77E-16&    0.081&  $<$5.67E-15&      \nodata&     0.11\\
IC348-67 &2.40E-15&  6.69E-16&    0.018&  $<$2.73E-15&      \nodata&    0.019&  $<$3.43E-15&      \nodata&    0.023\\
IC348-72&  $<$2.93E-15&      \nodata&    0.011&  $<$2.04E-15&      \nodata&    0.011&  $<$2.12E-15&      \nodata&    0.012\\
IC348-133&  $<$2.90E-15&      \nodata&    0.017&  $<$2.69E-15&      \nodata&    0.016&  $<$4.85E-15&      \nodata&    0.023\\
DM~Tau & 5.45E-15&  4.59E-16&    0.047&  $<$2.13E-15&      \nodata&    0.048&  $<$1.56E-15&      \nodata&    0.055\\
FP~Tau	& $<$4.73E-15&      \nodata&    0.067&  $<$3.76E-15&      \nodata&    0.070&  $<$4.31E-15&      \nodata&    0.078\\
GM~Aur 	& 1.02E-14&  7.11E-16&    0.097&  1.89E-14&  9.28E-16&     0.10&  1.20E-14&  1.02E-15&     0.17\\
LkCa~15 	&  3.41E-15&  5.54E-16&     0.14&  2.94E-15&  6.71E-16&     0.15&  $<$7.89E-15&      \nodata&     0.27\\
St~34 	& 1.40E-15&  5.66E-16&    0.029&  1.59E-15&  4.68E-16&    0.028&  $<$2.12E-15&      \nodata&    0.029\\
UX~TauA 	&  9.04E-15&  9.85E-16&     0.11&  9.02E-15&  8.53E-16&     0.11&  $<$4.98E-15&      \nodata&     0.14\\
V836Tau&  $<$2.05E-15&      \nodata&     0.10&  $<$3.61E-15&      \nodata&     0.11&  $<$3.33E-15&      \nodata&     0.17\\
LkHa~330 	& $<$4.99E-15&      \nodata&     0.49&  8.84E-15&  1.91E-15&     0.50&  $<$2.49E-14&      \nodata&     0.69\\
2071-H9/ NGC2068 &1.21E-15&  3.75E-16&    0.023&  2.44E-15&  3.84E-16&    0.024&  $<$2.06E-15&      \nodata&    0.038\\
CHX 22 (T54)	&  5.48E-15&  9.30E-16&    0.032&  $<$2.57E-15&      \nodata&    0.036&  $<$1.04E-15&      \nodata&    0.049\\
CS~Cha  &	3.61E-14&  6.12E-16&    0.046&  6.11E-15&  2.53E-16&    0.048&  3.07E-15&  7.43E-16&    0.064\\
 SZ~Cha 	& 1.48E-14&  1.02E-15&     0.16&  6.10E-15&  1.32E-15&     0.17&  $<$6.93E-15&      \nodata&     0.25\\
SZ~18 	&  7.63E-16&  2.33E-16&   0.0075&  8.68E-16&  2.14E-16&   0.0073&  $<$2.63E-15&      \nodata&   0.0080\\
Sz~27 (T35) & 6.32E-15&  9.03E-16&    0.021&  3.03E-15&  7.27E-16&    0.022&  $<$2.67E-15&      \nodata&    0.026\\
TW~Hya & 5.01E-14&  2.07E-15&     0.61&  7.93E-14&  2.00E-15&     0.62&  3.34E-14&  3.64E-15&     0.81\\
IM~Lup 	&  7.03E-15&  1.54E-15&     0.48&  $<$4.58E-15&      \nodata&     0.49&  $<$6.37E-15&      \nodata&     0.60\\
DoAr~21 	&  7.80E-14&  2.58E-15&     0.65&  $<$1.32E-14&      \nodata&     0.63&  $<$4.66E-14&      \nodata&     0.86\\
 \hline
\multicolumn{10}{c}{\emph{Debris disks}} \\
\hline
1RXS~J121236.4-55203&  $<$6.09E-16&      \nodata&    0.014&  $<$8.71E-16&      \nodata&    0.016&  $<$8.11E-16&      \nodata&    0.019\\
1RXS~J122233.4-533347 & $<$4.75E-16&      \nodata&    0.020&  8.56E-16&  1.95E-16&    0.021&  $<$7.81E-16&      \nodata&    0.026\\
1RXS~J130153.7-53044&  $<$4.95E-16&      \nodata&   0.0084&  $<$4.67E-16&      \nodata&   0.0092&  $<$8.19E-16&      \nodata&    0.011\\
 HD119269&  $<$1.15E-15&      \nodata&    0.024&  $<$9.33E-16&      \nodata&    0.026&  $<$1.49E-15&      \nodata&    0.031\\
1RXS~J133758.0-413448 & $<$6.79E-16&      \nodata&    0.019&  5.44E-16&  2.04E-16&    0.021&  $<$8.21E-16&      \nodata&    0.025\\
1RXS~J161458.4-27501&  $<$8.43E-16&      \nodata&   0.0088&  $<$6.31E-16&      \nodata&   0.0098&  $<$8.00E-16&      \nodata&    0.012\\
HD~142361 	& $<$1.59E-15&      \nodata&    0.040&  2.98E-15&  1.47E-16&    0.043&  1.44E-15&  2.60E-16&    0.052\\
RX~1600.0-2159 & $<$7.96E-16&      \nodata&    0.012&  1.20E-15&  1.91E-16&    0.013&  $<$7.63E-16&      \nodata&    0.015\\
RX~1612.2-1859 & $<$6.28E-16&      \nodata&    0.028&  9.32E-16&  1.53E-16&    0.031&  $<$8.60E-16&      \nodata&    0.036\\
HD~35850 &3.38E-15&  1.01E-15&     0.26&  9.02E-15&  9.27E-16&     0.27&  $<$6.30E-15&      \nodata&     0.34\\
V343Nor 	& $<$3.30E-15&      \nodata&     0.11&  4.81E-15&  8.09E-16&     0.12&  $<$5.64E-15&      \nodata&     0.15\\
HD19668&  $<$1.14E-15&      \nodata&    0.051&  $<$1.62E-15&      \nodata&    0.056&  $<$2.56E-15&      \nodata&    0.067\\
HD25457&  $<$1.86E-14&      \nodata&     0.53&  $<$1.53E-14&      \nodata&     0.58&  $<$1.86E-14&      \nodata&     0.68\\
HD37484&  $<$2.66E-15&      \nodata&    0.080&  $<$2.16E-15&      \nodata&    0.086&  $<$4.47E-15&      \nodata&     0.10\\
HD12039&  $<$7.55E-16&      \nodata&    0.065&  $<$1.31E-15&      \nodata&    0.071&  $<$1.43E-15&      \nodata&    0.082\\
HD202917&  $<$1.94E-15&      \nodata&    0.044&  $<$3.19E-15&      \nodata&    0.048&  $<$4.31E-15&      \nodata&    0.058\\
HD~141943	&$<$1.27E-15&      \nodata&    0.074&  3.10E-15&  4.13E-16&    0.080&  2.97E-15&  4.91E-16&    0.094\\
 HD377&  $<$1.79E-15&      \nodata&    0.088&  $<$1.31E-15&      \nodata&    0.095&  $<$2.73E-15&      \nodata&     0.11\\
HD17925&  $<$3.42E-15&      \nodata&     0.61&  $<$7.36E-15&      \nodata&     0.66&  $<$8.26E-15&      \nodata&     0.78\\
HD32297&  $<$3.26E-15&      \nodata&    0.054&  $<$4.28E-15&      \nodata&    0.053&  $<$5.09E-15&      \nodata&    0.052\\
HD61005&  $<$1.49E-15&      \nodata&    0.066&  $<$1.32E-15&      \nodata&    0.072&  $<$2.40E-15&      \nodata&    0.084\\
HD69830&  $<$4.54E-15&      \nodata&     0.56&  $<$4.75E-15&      \nodata&     0.62&  $<$1.05E-14&      \nodata&     0.81\\
HD134319&  $<$2.21E-15&      \nodata&    0.047&  $<$2.55E-15&      \nodata&    0.051&  $<$2.73E-15&      \nodata&    0.060\\
 HD216803&  $<$5.66E-15&      \nodata&     0.74&  $<$7.99E-15&      \nodata&     0.81&  $<$6.96E-15&      \nodata&     0.95\\
AOMen&  $<$1.04E-15&      \nodata&    0.051&  $<$1.03E-15&      \nodata&    0.056&  $<$2.08E-15&      \nodata&    0.065\\

\enddata
\tablecomments{$^a$Fluxes and errors are in erg s$^{-1}$ cm$^{-2}$. For the detections we report the 1$\sigma$error, upper limits are at 3$\sigma$. $^b$ Continua are in Jy. Objects marked with an $^*$ refer to water-contaminated disks, and the fluxes reported are already corrected for the water contribution. In parenthesis we report the observed {\hi} fluxes before the water subtraction. }
\end{deluxetable*}

\end{document}